\documentclass[graybox, nosecnum]{svmult}

\usepackage{mathptmx}       
\usepackage{helvet}         
\usepackage{courier}        
\usepackage{type1cm}        
\usepackage{makeidx}         
\usepackage{graphicx}        
\usepackage{multicol}        
\usepackage[bottom]{footmisc}
\usepackage{hyperref}        
\usepackage{soul}            

\usepackage{amsmath,amssymb} 

\usepackage{aas_macros} 
\usepackage{natbib} 
\usepackage{color}

\newcommand \beq{\begin{eqnarray}}
\newcommand \eeq{\end{eqnarray}}
\newcommand{\rmd}{\mathrm{d}}

\newcommand{\rme}{\mathrm{e}}

\makeindex             

\begin{document}
\title*{Equation of state in neutron stars and supernovae}
\author{Kohsuke Sumiyoshi \thanks{corresponding author}, Toru Kojo, and Shun Furusawa }
\institute{
Kohsuke Sumiyoshi \at National Institute of Technology, Numazu College, Numazu, Shizuoka 410-8501, Japan, \email{sumi@numazu-ct.ac.jp}
\and
Toru Kojo \at Department of Physics, Tohoku University, Sendai 980-8578, Japan \email{toru.kojo.b1@tohoku.ac.jp}
\and Shun Furusawa \at College of Science and Engineering, Kanto Gakuin University, Kanazawa-ku, Yokohama, Kanagawa 236-8501, Japan  \email{furusawa@kanto-gakuin.ac.jp},  \at Interdisciplinary Theoretical and Mathematical Sciences Program (iTHEMS), RIKEN, Wako, Saitama 351-0198, Japan
}
%
%
\maketitle
\abstract{
Neutron stars and supernovae provide cosmic laboratories of highly compressed matter at supra nuclear saturation density which is beyond the reach of terrestrial experiments.  
The properties of dense matter is extracted by combining the knowledge of nuclear experiments and astrophysical observations via theoretical frameworks. 
A matter in neutron stars is neutron rich, and may further accommodate non-nucleonic degrees of freedom such as hyperons and quarks. 
The structure and composition of neutron stars are determined by equations of state of matter, which are the primary subject in this chapter.
In case of supernovae, the time evolution includes several dynamical stages whose descriptions require equations of state at finite temperature and various lepton fractions.
Equations of state also play essential roles in neutron star mergers which allow us to explore new conditions of matter not achievable in static neutron stars and supernovae.
Several types of hadron-to-quark transitions, from first order transitions to crossover, are reviewed, and their characteristics are summarized.
}

\section{\textit{Introduction: Matter in the Cosmos}}

Exploration to the world at extreme conditions is one of the most fascinating themes in science.  
Expedition to the high density and temperature of matter in the Universe goes far beyond the experimental ranges attained on the Earth.  
Neutron stars and supernovae are such cosmic laboratories where new phases of matter, including hyperons and quarks, may be realized.  
It is thrilling, at the same time, to envisage the exotic matter from the information of nuclei at the limiting condition.  
Experimental studies of exotic nuclei are powerful tools to extend the knowledge on the hot and dense matter with theoretical models.  
This endeavor has been made for decades and the expedition is rapidly advancing the frontier with modern technology of nuclear experiments and astrophysical observations.

The purpose of this chapter is to provide the basic knowledge of hot and dense matter in compact objects, i.e., neutron stars, supernovae, and neutron star mergers in the Universe, 
and is to delineate the relation between these astrophysical objects and the properties of dense matter in quantum chromodynamics (QCD).
Our discussion starts with the overview of the compact objects and proceeds to the examination of conditions such as density, temperature, and composition, realized in the compact objects. 
These variables specify equations of state of matter which play crucial roles in determining the structural and dynamical aspects of neutron stars.
The conditions realized in supernovae or neutron star mergers are reviewed.
A matter in the similar conditions may be also studied by laboratory experiments, i.e., heavy ion collisions and unstable nuclei, 
and such similarity encourages the interplay between nuclear physics and astrophysics.
Near the nuclear saturation density, nuclear many-body theories
give important constraints on the structure of neutron stars, and they have been implemented in the analyses of neutron star observables.
Heavy neutron stars may accommodate matters at densities several times greater than the saturation density,
where the appearance of hyperons and quarks may change the overall trend of nuclear equations of state.
The latter part of this chapter examines the characteristics of nuclear and quark matters and then classifies
several types of hadron-to-quark matter transitions.
In this review, the natural unit, $c=\hslash=1$, is used unless otherwise stated.

\section{\textit{Properties of neutron stars and supernovae}}

Neutron stars are highly compact objects with a mass of $1-2M_{\odot}$ ($M_\odot$: solar mass) and a radius of $\sim$10 km, which means the average 
mass density, $m_N n$ ($m_N \simeq 939$ MeV: nucleon mass), of $\sim7\times10^{14}$~g/cm$^3$ \citep{Shapiro_Book}.  
This density is about twice as high as the nuclear density $\rho_0= m_N n_0 = 3\times10^{14}$~g/cm$^3$ ($n_0=$0.16~fm$^{-3}$)
for a canonical neutron star with the mass $\sim 1.4M_\odot$, and is even higher for more massive neutron stars with $\sim 2M_\odot$, 
possibly including a matter beyond the purely nucleonic regime.
The mass and radius of a neutron star are determined by equations of state which reflects the properties of strongly correlated matter.
The competition between pressure and energy density of a matter is essential;
the energy density induces the gravitational attraction toward the center while pressure increase in such compression prevents the matter from collapsing.
If the energy density dominates over pressure, massive matter collapses to a black hole. 
A matter having a large (small) $P(\rho)$ at a given energy density is called {\it stiff (soft)},
and stiff equations of state allow the existence of very massive neutron stars.
The mass threshold dividing neutron stars and black holes, and
how stiff a neutron star matter can be, are key questions in this chapter.

\begin{figure*}[ht]
\centering
\includegraphics[width=0.99\textwidth]{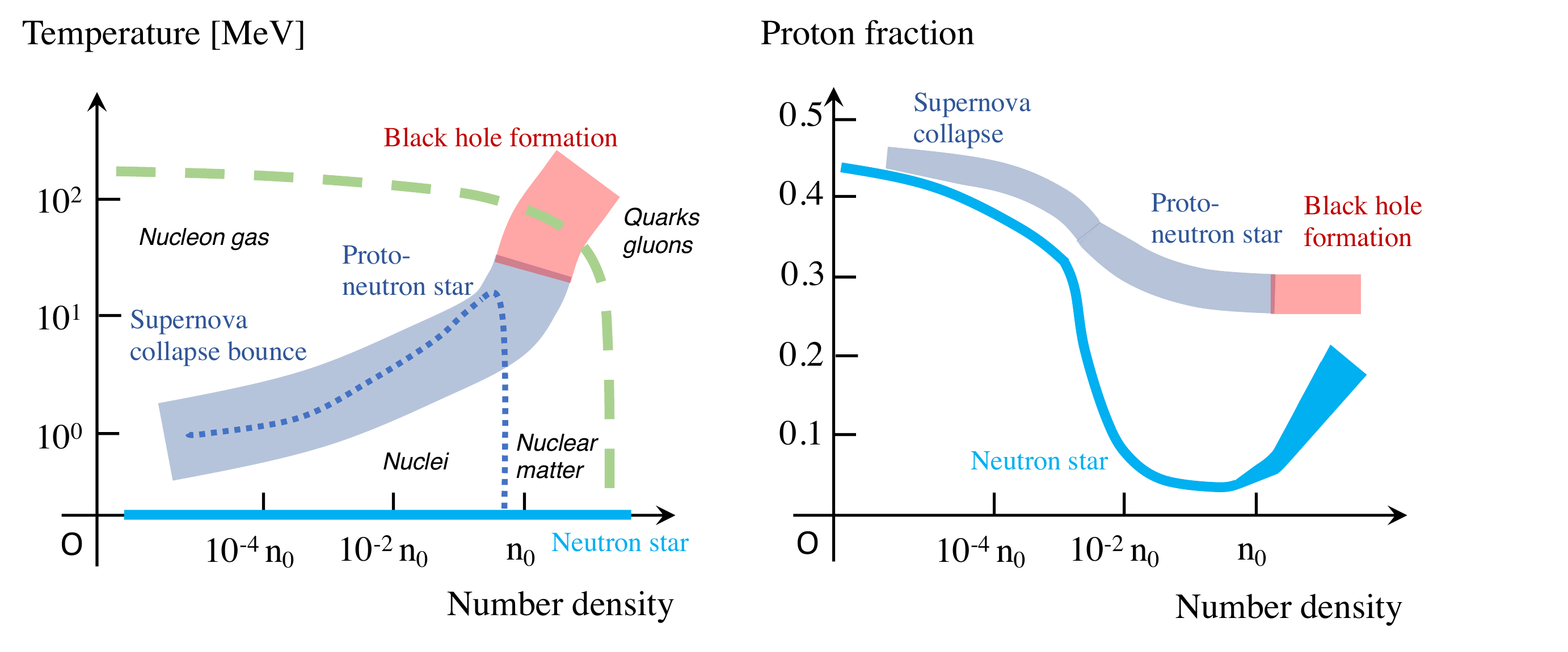}
\caption{Typical environment of neutron stars and supernovae is schematically shown in the plane of number density-temperature and number density-proton fraction.  Light blue thick lines show the conditions of neutron stars from the surface to the center.  The temperature of {\it cold} neutron stars is essentially zero.  The proton fraction of neutron star decreases as the density increases and becomes very small (neutron-rich) around $n_0$, but at higher density increases again due to the growth of the symmetry energy.  Shaded areas show the evolving conditions at the central region in the supernova core during the dynamics of collapse, bounce and the birth of proto-neutron star.  Red shaded areas show the conditions in the case of black hole formation due to the failure of explosion.  The phase boundaries for nuclei, nucleons and quarks-gluons are sketched by dotted and dashed lines. 
\label{fig:Intro_phase}}
\end{figure*}

As the name suggests, the interior of neutron stars is neutron rich; the proton fraction in nucleon density, $Y_p = n_p/n$, is $\sim 0.1$ at $n\sim n_0$,
and the positive charges are neutralized by charged leptons \citep{Shapiro_Book}.
The temperature of interior can be inferred to be $\sim10^{8}$~K ($\sim 0.01-0.1$~MeV), which is much smaller than the Fermi momenta of nucleons $ p_F \sim 300$ MeV.
In this sense neutron stars are regarded as cold.
In Fig. \ref{fig:Intro_phase}, typical environment is schematically shown in the phase diagram.  
This highly neutron rich and cold dense matter at $n > n_0$ is not achieved in terrestrial experiments such as heavy ion collisions
where the matter is isospin symmetric and inevitably accompany heat which results in temperature of  $10^{11}-10^{12}$~K ($\sim10-100$~MeV).
In this respect neutron stars are quite unique.

\begin{figure*}[ht]
\centering
\includegraphics[width=0.9\textwidth]{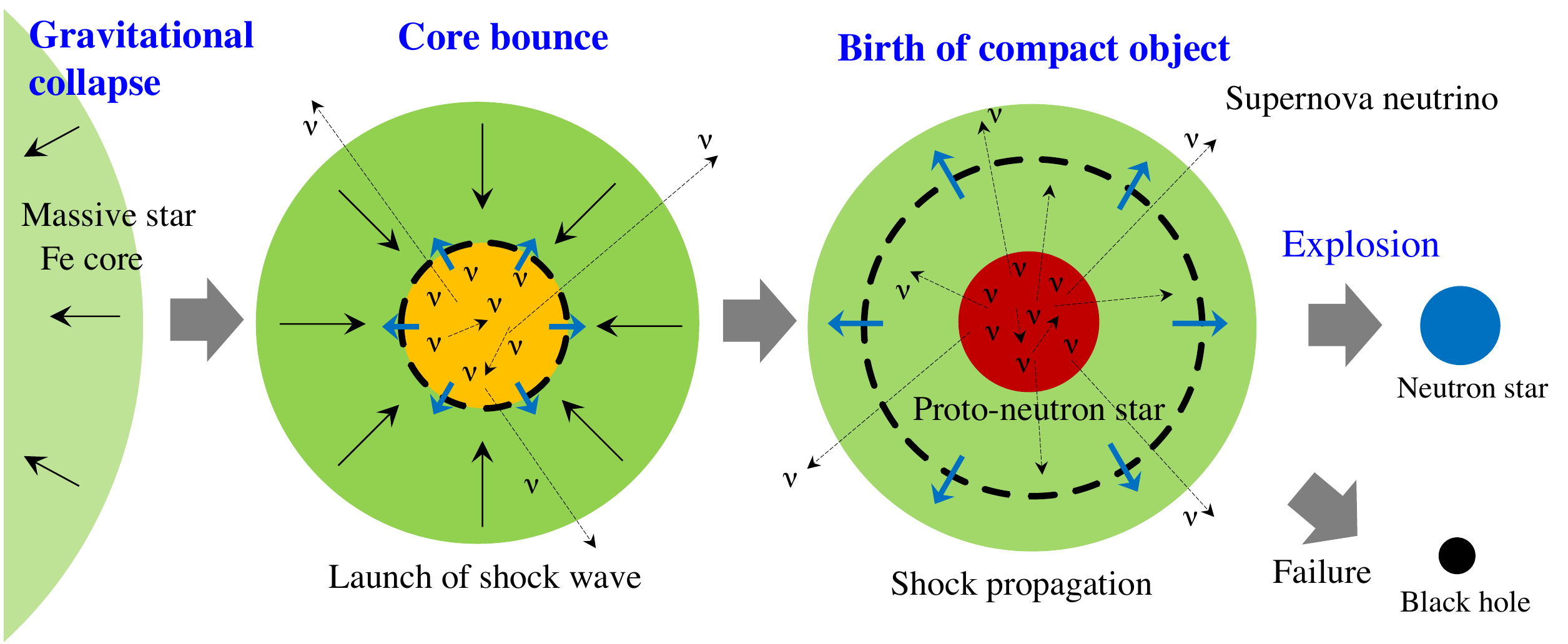}
\caption{Schematic diagrams of the evolution of supernova cores from massive stars to compact objects.  
Starting with the gravitational collapse of the Fe core, the central core is compressed 
and neutrinos are trapped inside due to interactions with hot and density matter.
The central core bounces back by a matter around the nuclear saturation density and the shock wave is launched.  
A proto-neutron star, which is hot and contains abundant neutrinos, is born at the center.  
If the shock wave successfully propagates through accreting outer layers of the Fe core, it leads to the supernova explosion and the birth of neutron star.  
If it fails, the central object collapses to the black hole.  
\label{fig:Intro_supernova}}
\end{figure*}

Meanwhile in supernovae that give the birth of neutron stars, the conditions similar to heavy ion experiments are realized \citep{Hempel2015,Oertel:2016bki}.
Supernova explosions occur at the end of stellar life after $\sim 10^7$~years, 
leaving the compact objects (neutron stars and black holes) as remnants at the center as shown in Fig. \ref{fig:Intro_supernova}.
As a result of gravitational collapse of the massive star, a proto-neutron star is born after bounce back of the central core just above the nuclear density.  
The success or failure of the propagation of shock wave, which is launched by the core bounce, 
determines the fate of the massive star; the supernova explosion leaving a neutron star or the black hole formation.  
During supernovae explosion, a hot and dense matter from dilute to dense conditions with the number density ranging from $10^{-10}$ to $10^{0}$~fm$^{-3}$,
the temperature of $T= 1-50$ MeV, and the proton fraction (proton density per baryon density) $Y_p = 0.3-0.4$, appears in the dynamical process as shown in Fig. \ref{fig:Intro_phase}.  
Note that the matter is not yet fully neutron-rich in the proto-neutron star due to the existence of neutrinos.  
If the collapse to a black hole happens, the density and temperature may become higher 
and equations of state with even broader range of ($n, Y_p, T$) contributes.

In the context of matter in quantum chromodynamics (QCD), cold neutron stars and supernovae offer information for the {\it QCD phase diagram}
expanded in a density (chemical potential) - temperature plane.
The astrophysical observations cover the domain not explored by terrestrial experiments 
and {\it ab-ininio} lattice QCD Monte-Carlo simulations.
The lattice QCD has been very powerful tools to study the QCD vacuum and the high temperature domain, but its application to finite density is prevented by the sign problem.

\section{\textit{Observations and the properties of matter}}

It should be useful to glance at how the above mentioned properties have been inferred from observations \citep{Shapiro_Book,Haensel2007}. 
Below some basic observations are mentioned.
Neutron stars can contain large magnetic fields and emit strong pulses of electromagnetic waves. 
Such pulse emitters are called pulsars.
The pulses arriving at the earth is periodic as neutron stars are rotating.
The extremely precise and stable frequency of signal with a very short period of $\sim$1 ms indicates that the rotating object is very compact;
otherwise an object with the large radius $R$ and high frequency $\omega$ would lead to the velocity of $\sim \omega R$ exceeding the light velocity.
There are many pulsars which are discovered in a binary composed of a neutron star and another star which can be a neutron star or a white dwarf or a black hole \citep{Lorimer2008}. 
Such systems are useful to estimate the neutron star mass, as the Kepler's third law can be applied to the binary motion \cite{Lattimer2012,Oezel2012,Kiziltan2013}.
The most precise mass measurement further utilizes the general relativistic time delay on pulses passing gravitational fields around the companion star \citep{Demorest2010,Antoniadis2013}.
The heaviest neutron star discovered is PSR J0740+6620 with the mass $2.08\pm 0.07M_\odot$ \citep{Cromartie2020,Fonseca2021}
which put tight constraints on the lower bound of the stiffness of equations of state.

The measurements on neutron star radii are more difficult than the mass measurements \citep{Fortin2015,Oezel2016}. 
 The analyses of X-ray bursts on the neutron star surface lead to the estimate of $\sim$10 km, 
 but they contain several uncertainties such as modeling and the distance between the earth and neutron stars \citep{Watts:2016uzu}.
In this respect, dramatic progress took place from the first detection of gravitational waves from the neutron star merger event, GW170817 \citep{abb18}.
The gravitational wave patterns are very sensitive to the (tidal) deformation of each neutron star just before the collision, 
since such deformation induces additional gravitational attraction between neutron stars, accelerating the merging process.
The tidal deformation strongly depends on the neutron star radii; a larger radius leads to a larger tidal deformability. 
For a $1.4M_\odot$ neutron star the upperbound is $R < 13.4$ km \citep{Annala2018,Most:2018hfd}.
The LIGO collaboration yielded the estimate $R =11.9^{+1.4}_{-1.4}$ for neutron stars in GW170817 \citep{abb18}.
Another important constraint was set by the NICER measurement of X-rays from hotspots of rotating neutron star surfaces \citep{mcmil19,ril21,mcmil21}.
It keeps track of the hotspots and examine the Doppler shift of the X-ray spectra from moving hotspots, taking gravitational lensing into account.
The measurements of the rotation frequency $\omega$, the surface velocity $\sim R \omega$, and the gravitational lensing related to $M/R$
in principle allow us to extract $M$ and $R$ separately.
The NICER analyses have been done for $1.4M_\odot$ and $\sim 2.1M_\odot$ neutron stars, and the two radii turn out to be rather similar, 
$12.45 \pm 0.65$ km for a $1.4 M_\odot$ neutron star and $12.35 \pm 0.75$ km for a $2.08 M_\odot$ neutron star \citep{mcmil21}.
%

The birth places of neutron stars are considered to be supernovae explosions \citep{bet90}. 
In fact pulsars are found in the supernova remnant as in the case of Crab pulsar in the remnant of supernova in 1054.  
Supernovae are transient phenomena with a bright display for months and disappearance afterward.  
A few events of supernovae occurs on the average per century in a galaxy, but more than 1000 cases are recorded per year in the monitoring surveys.  
The process of ending in the stellar life 
includes the matter evolution through nuclear fusion, dissociation, 
electron captures 
and subsequent neutrino production, as well as thermal production of neutrino-antineutrino pairs, affecting the composition of heavy elements \citep{Arnett1996}.  
From observation of light curves and spectra with modeling of explosive nucleosynthesis, the energy of explosion is estimated to be $\sim10^{51}$~erg.  In the case of supernova observed in 1987, neutrino bursts from supernova SN 1987A were detected with average energy $\sim$10~MeV for $\sim$10~s at the terrestrial detectors \citep{Hirata1987,Bionta1987}.  
The detection of supernova neutrinos vindicates the birth of neutron star, which has a condition at high density and temperature.  
The total energy of neutrinos is evaluated to be $\sim3\times10^{53}$erg \citep{KSato1987,Burrows1988,suz94}, which is well in accord with the gravitational binding energy of neutron stars,
supporting the scenario of supernova explosion driven by neutrinos.  
In fact, neutrinos play an essential role in the explosion mechanism and controls properties of hot and dense matter.  
At the terrestrial detectors such as Super-Kamiokande and IceCube, a burst of supernova are being monitored to find neutrinos from astrophysical phenomena.  
In the next nearby supernova,
the improvement of the detectors allows us to detect $\sim 1000$ times more neutrinos than found in SN 1987A \citep{Scholberg2012,hypk2018,suw19}.
The detailed information of the energy spectra of neutrinos as a function of time will be used to reveal the explosion dynamics and the properties of compact object.  
The supernova explosion is also a target of the observation of gravitational waves, 
which is generated by the time dependent quadrupole energy distribution and thus
carries the information of multi-dimensional dynamics, at the detectors such as LIGO, VIRGO and KAGRA \citep{Kotake2006,Ott2009,kot13}.  

These observations are rapidly progressing. 
The gravitational wave detection as well as the NICER measurements have just begun.
When the gravitational detector reaches the design sensitivity, the detection of gravitational waves from neutron star mergers should be daily events.
The NICER is collecting more photons from neutron stars and improving the statistics, and at the same time is increasing the number of targets.
Supernovae events in our galaxy may happen within several decades and should dramatically improve our understanding of supernova matter as well as the properties of neutrinos \citep{Ando2005}.
The diffuse supernova neutrino background, which has accumulated from the neutrino bursts from the collapse of massive stars, 
will be detected in the coming years \citep{Horiuchi2009}.
Our aim in the following sections is to summarize basic aspects of neutron star and supernova matter 
and to prepare ourselves for the expected and unexpected new discoveries.

\subsection{The structure of neutron star and equation of state}

\begin{figure*}[ht]
\vspace{-1.5cm}
\centering
\includegraphics[width=0.70\textwidth]{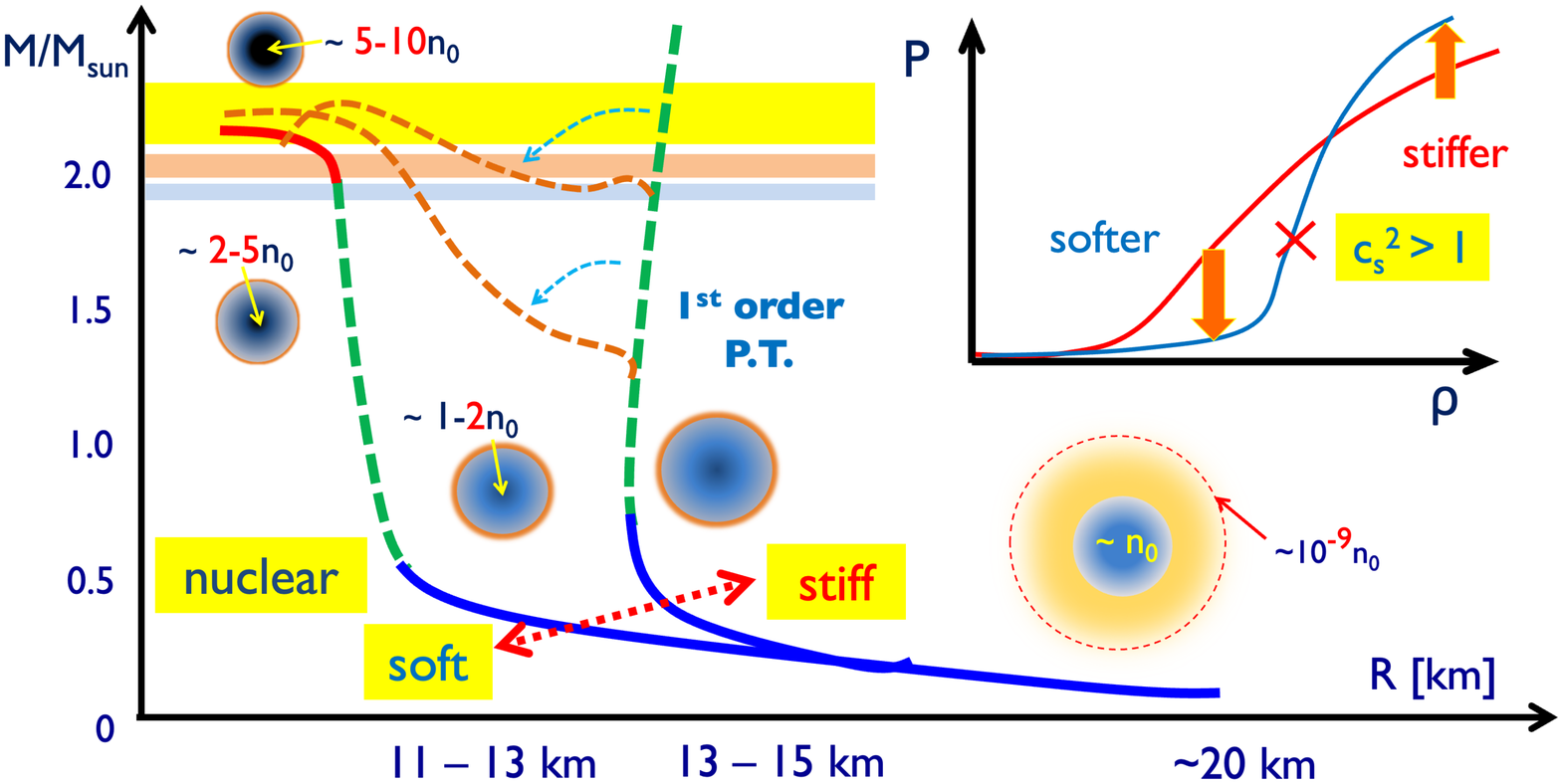}
\vspace{-0.8cm}
\caption{ $M$-$R$ curves as functions of the central density $n_c$.
For a small $M$ neutron stars have large radii due to loosely bound crust.
When the central density reaches $n_c = 1\sim 2n_0$, the dilute matter is highly compressed and the size of a neutron star 
is characterized by a matter beyond the saturation point.
The curves go up with small variation in the radii. 
The exception is equations of state with the first order transitions which lead to kinks in the $M$-$R$ curves.
The top right figure illustrates how the causality constrains the relation between low and high density behaviors of $P(\rho)$.
\label{fig:M-R}}
\end{figure*}

The most basic quantities in neutron star observations are the mass-radius ($M$-$R$) relations of neutron stars.
As mentioned, the structure is determined by pressure v.s. energy density, $P(\rho)$, which enters {\it Tolman-Oppenheimer-Volkoff} (TOV) equation,
a general relativistic version of the Newton equation for gravity, 
\begin{equation}
    \frac{dP(r)}{dr}=-\frac{GM(r)\rho(r)}{r^2} \bigg( 1 + \frac{\, P \,}{\rho} \bigg) \bigg( 1 + \frac{\, 4\pi r^3 P \,}{\, M(r) \,} \bigg) \bigg( 1 - \frac{\, 2G M \,}{r} \bigg)^{-1} \,.
\label{eqn:NSeq1}
\end{equation}
where the last three factors characterize the general relativistic effects, $P(r)$ and $\rho(r)$ are the pressure and density at the radial position, $r$ \citep{Shapiro_Book}.  
The mass $M(r)$ inside of the radial coordinate $r$ is obtained by integrating the mass shell at $r$,
\begin{equation}
    \frac{dM(r)}{dr}=4 \pi r^2 \rho(r) \,.
\label{eqn:NSeq2}
\end{equation}
Actual computations begin with setting the central density $n_c$ at $r=0$ and then an equation of state is used to prepare $P$ and $\rho$ at given $n_c$.
With these initial conditions and equations of state $P(\rho)$, the differential equations \ref{eqn:NSeq1} and \ref{eqn:NSeq2} are integrated until the pressure reaches zero, $P(r=R)=0$,
which defines the radius $R$ of a neutron star. The neutron star mass is defined as $M(r=R)$.
In this way, for a given central density $n_c$, one can obtain $M(R; n_c)$. 
Repeating this procedure with changing $n_c$,
one obtains the collection of such data forming a $M$-$R$ curve. 
There is one-to-one correspondence between $M$-$R$ curves and equations of state;
in principle the equation of state can be directly extracted from observations.
In reality there are only a few examples where $M$ and $R$ are determined simultaneously.
In this situation some guides from theories should be supplemented.

From the center to the surface, the density drastically drops from $\sim 2-5n_0$ to $\sim 10^{-10} n_0$, 
and apparently it seems a formidable task to extract equations of state from such complex structure.
Actually details of dilute matter do not have significant impacts on the $M$-$R$ relations, 
since dilute matter does not have large energy and its domain is highly compressed by the gravity;
dilute matter with $n\ll n_0$ forms only a thin shell of $\sim 0.5$ km at most.
A possible exception to this discussion is very light neutron stars with $M \sim 0.1-0.5M_\odot$ for which dilute matter is loosely compressed, 
but such neutron stars have not been discovered perhaps due to the absence of the formation process for such light neutron stars.
With this considerations, one can concentrate on equations of state for $n \gtrsim 0.1n_0$ for discussions of $M$-$R$ curves.
(The dilute matter part is crucial to discuss phenomena near the neutron star surfaces or finite temperature
as the loosely bound matter is quite active.) \citep{Page2006,Haensel2007}

It is very important to note that the shapes of $M$-$R$ curves are strongly correlated with the stiffness at several fiducial densities (Fig. \ref{fig:M-R}).
Stiff (soft) equations of state lead to larger (smaller) radii, and larger (smaller) maximum masses.
From low mass to the mass $\sim 1.0M_\odot$, $M$-$R$ curves first show the shrinkage of $R$, 
and then radically changes the direction toward the vertical direction with small changes in $R$.
This bending occurs with the central density around $\sim n_0$. 
This reflects that the compressed matter begins to observe repulsive forces in nuclear matter and matter can no longer be squeezed easily.
The location of the bending sets the overall radius of neutron stars. 
Thus neutron star radii are typically sensitive to equations of state at $1-2n_0$ for wide range in $M$.
The exceptions to this rule are equations of state with first order phase transitions;
the associated radical softening induces kinks in $M$-$R$ curves,
although up to now the signature of such kink structure has not been found for the interval $1.4-2.1M_\odot$.
Beyond $\sim 1.4M_\odot$, the core density typically exceeds $2-3n_0$, entering the regime beyond purely nucleonic regime (see discussions below),
and around $\sim 2 M_\odot$, the density may reach $4-7n_0$.
The existence of $2M_\odot$ neutron stars requires the high density matter at $4-7n_0$ to be very stiff.

In addition to observational constraints on $M$ and $R$, there are constraints on the interplay between low and high density equations of state.
The first is the causality constraint, $d P/d \rho = c_s^2 \le 1$, which demands the sound velocity $c_s$ to be smaller than the light velocity.
Another constraint is the thermodynamics stability, $d^2 \rho/dn^2 \ge 0$.
For example, one cannot combine extremely soft low density equations of state and extremely stiff high density ones, 
since $dP/d\rho$ must grow too rapidly from low to high densities.
With these constraints, it becomes theoretically more challenging if the maximum masses of neutron stars are larger and the radii are smaller
than the currently available constraints.
The low and high density equations of state constrain each other.

The following section starts with discussions of nuclear matter which have been studied intensively.
Matter in neutron stars and supernovae are not isospin symmetric, 
so it is necessary to discuss equations of state as functions of density $n$ and proton fraction $Y_p$.
The effects of temperatures are discussed for applications to supernova matter.
These discussions are given for neutron stars with the masses up to $\sim 1.4M_\odot$ and the core densities up to $2-3n_0$.
For $2M_\odot$ neutron stars, the core density is higher and more hypothetical arguments are needed.
Discussions related to quark matter is postponed to the final part of this review.

\section{\textit{Basic properties of dense matter}}

\subsection{ideal Fermi gas}

In order to understand the conditions in astrophysical phenomena, it is helpful to recall the basic properties of ideal gas of electrons and nucleons \citep{Lang,Shapiro_Book,Cox_Giuli}.  
Simple relations of the thermodynamics for the degenerate gas of fermions provide the energy scale in neutron stars and nuclei.  
The number density of fermion gas is given by $n=g p_F^3/3\pi^2$ where $p_F$ is the Fermi momentum and $g$ is the number of degree of freedom for spin.  
For example, at the initial stage of supernovae, electrons in Fe cores with the number density $n$ and $Y_p = 0.46$ have $p_F=10$~MeV $(n/10^{-5}~{\rm fm}^{-3})^\frac{1}{3}$
which is much greater than the electron mass $m_e \simeq 0.5$ MeV.
Another characteristic scale is the nucleon Fermi momenta in symmetric matter ($Y_p=0.5$) which is evaluated to be $p_F=263$~MeV $(n/n_0)^\frac{1}{3}$.  
The nucleons become relativistic when $p_F \simeq m_N$ or $n\simeq 50n_0$.

For a gas in the relativistic limit, the energy density scales as $\rho \propto p_F^4 \propto n^{4/3}$.
Using the relation $\mu = \partial \rho /\partial n$ ($\mu$: chemical potential) and the thermodynamic relation $P=\mu n- \rho$, one can write
\beq
\rho_{\rm rela} (n) = c n^{4/3} ~~\rightarrow ~~ P_{\rm rela} = \rho_{\rm rela}/3 \,.
\eeq
This regime is relevant for electrons in Fe cores or neutron stars. 
Meanwhile, if fermions with the masses $m$ are non-relativistic ($m\gg p_F$), 
the energy density behaves as $\rho \sim c_1 m n + c_2 n^{5/3}/m$ with $c_1,c_2$ being some constants.
In this case
\beq
\rho_{\rm NR} (n) = c_1 m n + c_2 \frac{\, n^{5/3} \,}{\, m \,}
~~\rightarrow~~ 
P_{\rm NR} = \frac{\, 2 \,}{\, 3 \,} c_2 \frac{\, n^{5/3} \,}{\, m \,} ~\propto~ \rho_{\rm NR}^{5/3} m^{-8/3} \,.
\eeq
It is to be noted that the energy density $mn$ is much larger than the pressure; a non-relativistic gas is soft.
In order for purely nucleonic descriptions to achieve a large neutron star mass of $\gtrsim M_\odot$, 
substantial repulsions among nucleons must be added to increase the pressure. 

In order to parameterize the effects of interactions and the evolution from non-relativistic to relativistic regimes of dense matter, one often uses a polytrope form,
$P=C \rho^{\Gamma}$, 
with $\Gamma$ called adiabatic index, and changes $\Gamma$ at some fiducial densities \citep{Read:2008iy}.
It has been known that a piecewide polytrope with the proper choice of $\Gamma$ and the fiducial densities can cover wide class of realistic equations of state.

\subsection{Nuclear matter}

The structure of neutron stars, especially the overall radii, is very sensitive to nuclear equations of state near the saturation density $n_0$ \citep{Lattimer:2000nx}.
Nuclear matter here is considered to be uniform and infinitely spread in space.
The saturation density $n_0$ is close to the density inside of heavy nuclei, and one can infer the properties of nuclear matter from laboratory experiments for finite size nuclei.
But the removal of finite size effects introduces uncertainties \citep{Atkinson:2020yyo}
 which would change the neutron star radii by $\sim 0.5-1$ km;
the pressure of nuclear matter is basically small, thus even small corrections from interactions affect the predictions for neutron stars.
Hence, the precise determination of nuclear matter properties still remains an important problem.
Below the basic properties of nuclear matter near saturation density  are summarized.
It is usually sufficient to evaluate contributions from nuclear interactions 
but neglecting the Coulomb energy among protons because the matter with leptons is locally charge neutral in most cases.
The interactions among leptons are negligible and leptons can be added separately as an ideal gas.

\begin{figure*}[ht]
\centering
\includegraphics[width=0.48\textwidth]{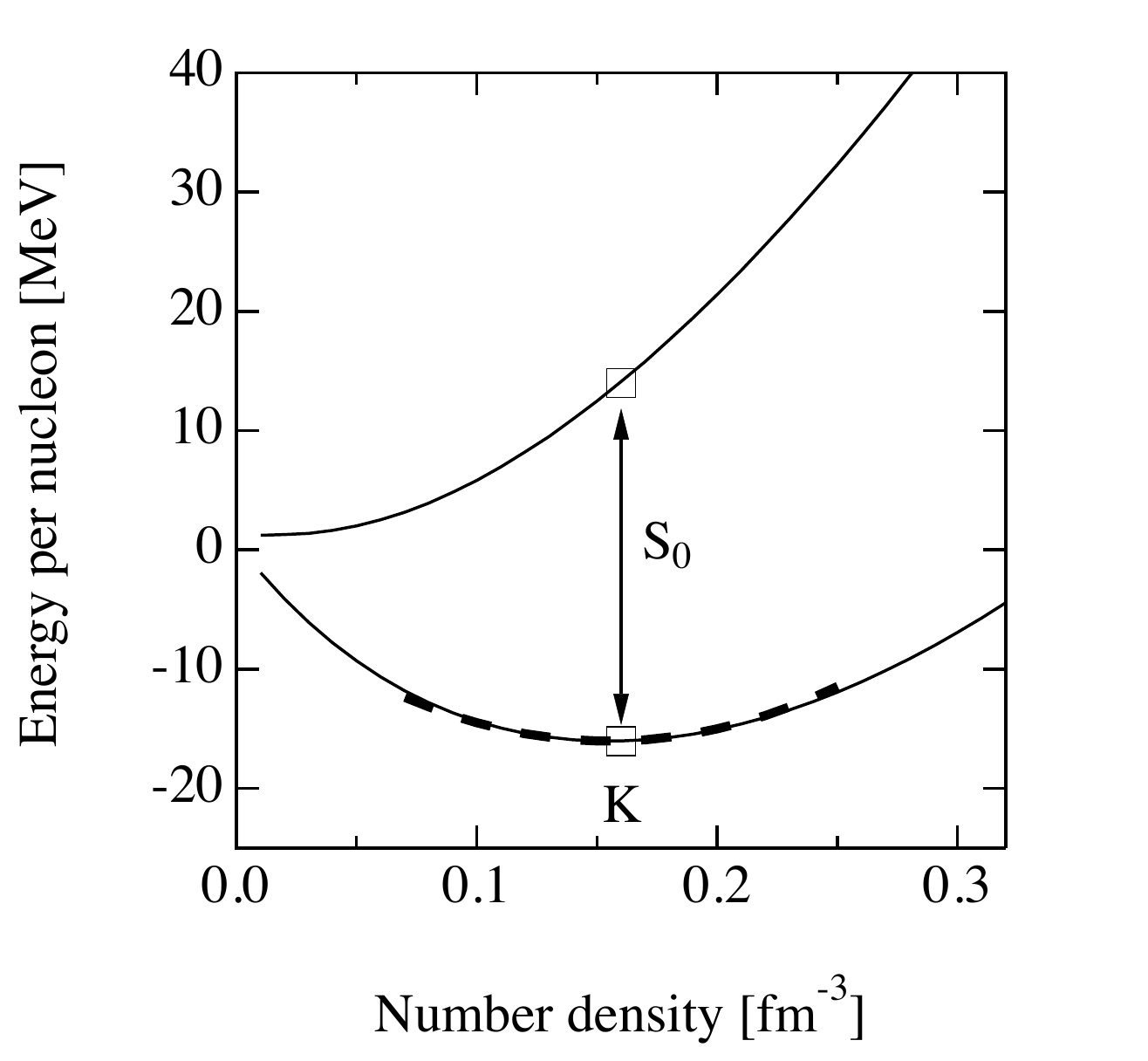}
\includegraphics[width=0.48\textwidth]{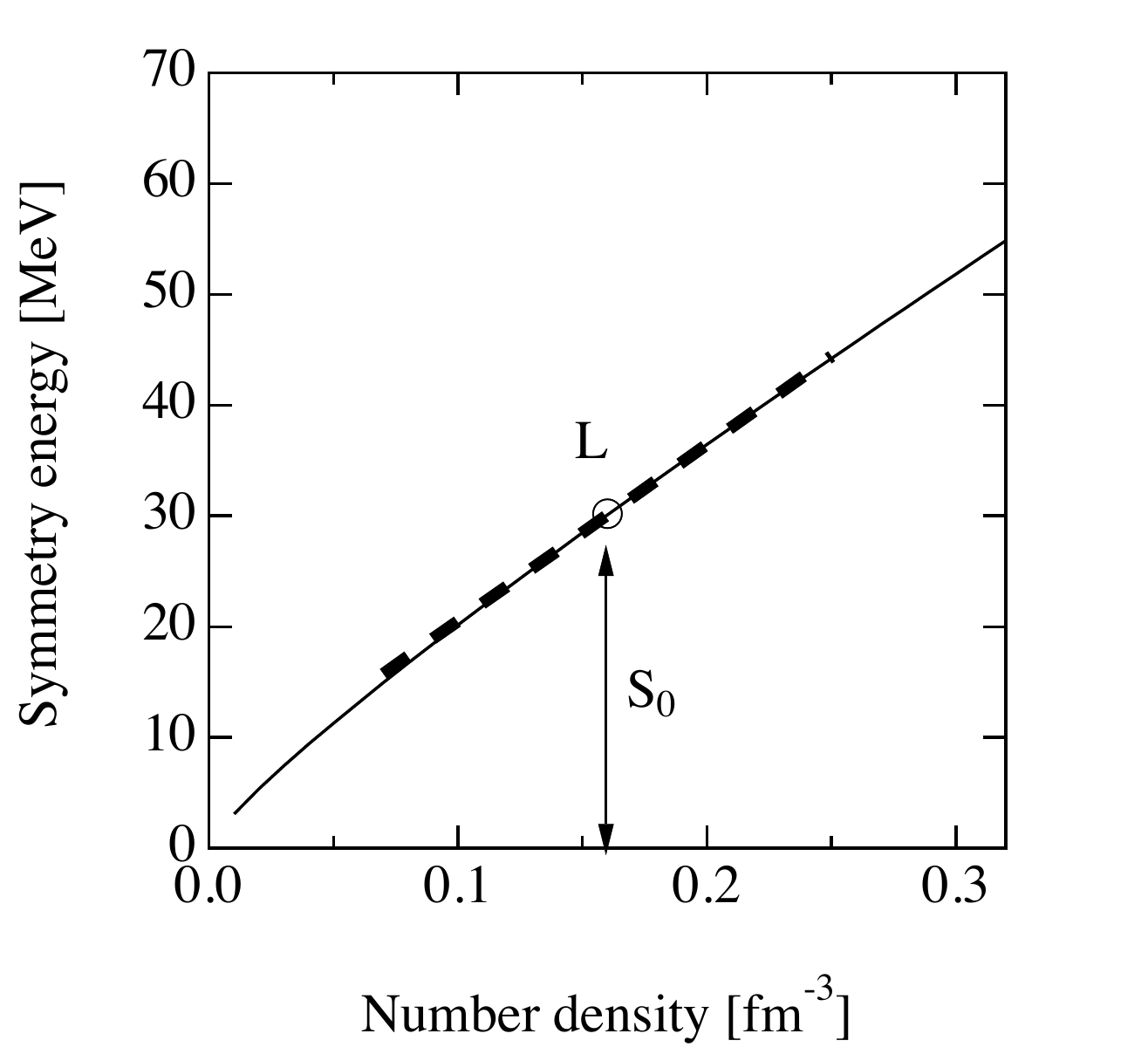}
\caption{Energy per nucleon of nuclear matter and neutron matter (left) and symmetry energy (right) is shown as a function of number density.  Approximate curves at the nuclear density are plotted by dashed lines with parabolic and linear expressions.  
\label{fig:EOS_nucl}}
\end{figure*}

Shown in Fig. \ref{fig:EOS_nucl} are typical behaviors of energies per nucleons with the mass subtracted, 
$\rho/n - m$, in symmetric matter ($Y_p=0.5$) and pure neutron matter ($Y_p=0$).
In symmetric matter, there is a minimum at $n_0=0.16$~fm$^{-3}$ with $\rho/n -m \simeq -16$~MeV. 
This is called saturation point.
Using the relation $P = \mu n- \rho = n^2 \partial (\rho/n)/\partial n$, one can conclude the pressure at $n_0$ is zero;
no external pressure is needed to maintain the finite matter, meaning that the matter is self-bound and stable against compression or expansion. 
This saturation property brings the stability of various nuclei at a constant density,
and the density greater than $n_0$ is achieved only by substantial external pressure. 
Meanwhile, pure neutron matter is not self-bound.  
Thus neutron stars are bound by the gravitational forces which require macroscopic amount of materials as the source.
Thus neutron stars cannot be arbitrarily light.

In applications to astrophysical phenomena, it is necessary to know the energy at various densities and proton fractions.  
The exploration to dense matter is often made by a simple form of the energy function, $E_N(n,Y_p) = \rho/n - m$, as
\begin{equation}
E_N(n,Y_p)=E_N(n_0,Y_p=0.5) + \frac{K(Y_p)}{18}\left(\frac{n -n_0}{n_0}\right)^2 +S(n)(1-2Y_p)^2 + \cdots
\end{equation}
in terms of the expansion around $n =n_0$ and $Y_p=0.5$.  
The coefficient $K(Y_p)$ is called the incompressibility, the curvature at $n_0$ for a given $Y_p$.
The coefficient $S(n)$ is called the symmetry energy which, in conventional literatures, 
is defined to be the second derivative of $E(n,Y_p)$ with respect to $Y_p$ at $Y_p=0.5$. 
This definition does not directly coincide with $E(n,Y_p=0) - E(n,Y_p=0.5)$,
but is chosen for the practical limitation that nuclear experiments cannot access the matter at $Y_p\sim 0$ where nuclei are unstable.
In practice, available theoretical calculations suggest keeping only the leading order of the expansion, 
which starts with $(1-2Y_p)^2$, is the good approximation around $n_0$. 
The symmetry energy near the nuclear density is expressed by $S(n) = S(n =n_0)+\frac{L}{3n_0}(n-n_0)$ using the slope parameter, $L$.  
These quantities are extracted to be $K(Y_p=0.5)=220-260$~MeV \citep{Shlomo2006,Stone2014,Garg2018}, 
$S(n = n_0)=24-36$~MeV, 
and 
$L=30-90$~MeV 
from the analyses of experimental data of nuclei for masses, radii and excitations \citep{tsa12,lat13,horo14,oze16,li18,li19}.  
These uncertainties are converted into the uncertainties in overall neutron star radii with $\Delta R \sim 0.5-2$ km.

\subsection{Nuclear matter theories}

The direct calculations of infinite nuclear matter have played important roles in the neutron star context \citep{Burgio2021}.
It can be applied to a matter with arbitrary $Y_p$, and 
especially pure neutron matter calculations are cleaner than symmetric matter due to fewer parameters in calculations.
The most systematic approach is based on microscopic two- and three-nucleon forces 
which are constrained by nuclear two-body scattering below the pion production threshold,
spectra of light nuclei, and a deuteron scattering off a proton that is sensitive to three-nucleon forces.
Either traditional potential models \citep{Stoks:1994wp,Machleidt:1995km}, which is based on the meson exchange picture, or chiral effective theory \citep{hol17,dri21}, 
which is based on momentum expansion, are used to characterize the nuclear forces \citep{Machleidt:2020vzm}.
These nuclear forces are then used in many-body framework such as variational \citep{Akmal:1998cf,Togashi:2017mjp}, 
quantum-Monte Carlo \citep{Carlson:2014vla}, or many-body perturbation theories \citep{Drischler:2017wtt}.
The modern calculations are consistent with the above-mentioned experimental estimates, 
but the pressure, which is sensitive to fine details, still has the uncertainties of $\sim 30$\% at $n_0$ (see, e.g., Fig.1 in \cite{Kojo:2021wax}).
These uncertainties grow with density, and
 in addition the validity to truncate many-body forces beyond three-body forces becomes questionable at $n \gtrsim 2n_0$.
 Thus a phenomenological modeling is often used beyond $n\sim 2n_0$ together with $2M_\odot$ constraints.
Taking the microscopic calculations as the low density constraints, typical radii of $1.4M_\odot$ neutron stars are $11.5-13$ km.

The thermodynamic conditions in core-collapse supernovae and neutron star mergers vary over a wide range of density, temperature, and proton fraction.  
To construct data of the equations of state  for the astrophysical simulations, phenomenological approaches are useful.
The Skyrme-type interactions, which are represented as expansions of the effective interaction
 in powers of momenta and density-dependent three-body contributions, 
 are the most well-known phenomenological modeling \citep{dut12}.
In the Skyrme-type models with the SLy4 parameters \citep{cha98}, 
the energy density of nuclear matter
is expressed as
\begin{eqnarray} 
\rho(n,Y_p,T)&=&\frac{ \tau_n}{2m_n^*}+
     \frac{ \tau_p}{2m_p^*}   +\left(a+4bx(1-x)\right) n^2 \nonumber  \\
& &  +\sum_{j=0,1,2} \left( c_j+4 d_j x(1-x)\right) n^{1+\delta_j} +(1- x) n m_n+ x n  m_p  \ ,    \label{eq:sky} \\
 \frac{1}{2m_{n}^*} &=&\frac{1}{2m_{n}}+\alpha_1 n_{n}+\alpha_{ 2n_{p} } \ ,  \label{eq_efm1} \\
 \frac{1}{2m_{p}^*}&=&\frac{1}{2m_{p}}+\alpha_1 n_{p}+\alpha_{ 2n_{n} }\  , \label{eq_efm2} 
\end{eqnarray}
where $a$, $b$, $c_j$, $d_j$, and $\delta_j$ are parameters of the
Skyrme forces, $\tau_n$ and $\tau_p$ are the kinetic energy
densities of neutrons and protons, respectively, and $m_n$ and $m_p$ are the masses of neutrons and protons, respectively.
 The first and second terms  in Eq.~(\ref{eq:sky}) correspond to the non-relativistic kinetic energy
densities of neutrons and protons, respectively.  The third term
 represents two-nucleon interactions,
 and  the summation over $j$  approximates the effects of many-body or density-dependent interactions.
The last two terms   in Eq.~(\ref{eq:sky}) express the rest masses of neutrons and protons, respectively.
The parameters $\alpha_1$ and $\alpha_2$  for effective masses, $m_n^*$ and $m_p^*$, are chosen to reproduce observables of uniform nuclear matter together with
$a$, $b$, $c_j$, $d_j$, and $\delta_j$.

Another phenomenological model of nuclear matter energy is the relativistic mean-field theory,
in which
nuclear interactions are described by the exchange of mesons. 
Up to this time, many parameter sets  in the relativistic mean field theory have been adopted to construct equations of state for astrophysical simulations, 
e.g., DD2 \citep{typ10}, SFHx, and SFHo \citep{ste13}.
They are subject to constraints from terrestrial experiments and astrophysical observations \citep{sto21}. 
For example, the Lagrangian  with a parameter set TM1e \citep{she20} is
($M_N$: nucleon mass)
\begin{eqnarray}
\label{eq:LRMF}
\mathcal{L}_{\rm{RMF}} & = & \sum_{i=p,n}\bar{\psi}_i
\left[ i\gamma_{\mu}\partial^{\mu}-\left(M_N + g_{\sigma}\sigma\right) \right. \nonumber \\
&& \left. -\gamma_{\mu} \left(g_{\omega}\omega^{\mu} +\frac{g_{\rho}}{2}
\tau_a\rho^{a\mu}\right)\right]\psi_i  \nonumber \\
&& +\frac{1}{2}\partial_{\mu}\sigma\partial^{\mu}\sigma
-\frac{1}{2}m^2_{\sigma}\sigma^2-\frac{1}{3}g_{2}\sigma^{3} -\frac{1}{4}g_{3}\sigma^{4}
\nonumber \\
&& -\frac{1}{4}W_{\mu\nu}W^{\mu\nu} +\frac{1}{2}m^2_{\omega}\omega_{\mu}\omega^{\mu}
+\frac{1}{4}c_{3}\left(\omega_{\mu}\omega^{\mu}\right)^2  \nonumber
\\
&& -\frac{1}{4}R^a_{\mu\nu}R^{a\mu\nu} +\frac{1}{2}m^2_{\rho}\rho^a_{\mu}\rho^{a\mu} \nonumber \\
&& +\Lambda_{\rm{v}} \left(g_{\omega}^2 \omega_{\mu}\omega^{\mu}\right)
\left(g_{\rho}^2\rho^a_{\mu}\rho^{a\mu}\right),
\end{eqnarray}
where $\psi$, $\sigma$, $\omega$, and $\rho$ denote nucleons, scalar-isoscalar mesons, vector-isoscalar mesons, and vector-isovector mesons, respectively, and 
$W_{\mu\nu} =\partial^{\mu}\omega^{\nu}- \partial^{\nu}\omega^{\mu} $ and 
$R^{a}_{\mu\nu}= \partial^{\mu}\rho^{a\nu}- \partial^{\nu}\rho^{a\mu} +g_{\rho}\epsilon^{abc}\rho^{b\mu}\rho^{c\nu} $. 
Nucleon-meson interactions are expressed as Yukawa couplings, and isoscalar mesons ($\sigma$ and 
$\omega$) interact with themselves. 
In the TM1e parameter set, the masses of mesons---$m_{\sigma}$, $m_{\omega}$, and $m_{\rho}$---and 
the coupling constants---$g_{\sigma}$, $g_{\omega}$, $g_{\rho}$, $g_2$, $g_3$,$c_3$, and $\Lambda_{\rm{v}}$---are 
determined to reproduce  both properties of 
uniform nuclear matter at the saturation density  \citep{oer17} and finite nuclei \citep{sug94,bao14a}. 
In the mean field theory, mesons are assumed 
to be classical and are replaced by their ensemble averages. 
The Dirac equations for nucleons are quantized, and the free energies are 
evaluated based on their energy spectra. 
The meson fields and Dirac equations are self consistently solved \citep{sum94}.

For astrophysical applications,
some 
 microscopic models have been constructed 
with realistic interactions determined using the nucleon--nucleon scattering data.
The variational method  \citep{tog13} for the Schrodinger's equation are based on the realistic two-body nuclear potentials, 
which are adjusted to account for the data, and on a three-body potential.
The Dirac Br\"{u}ckner Hartree--Fock  theory \citep{kat13} also employs the bare nuclear interactions. 
In contrast to variational method  with  a three-body potential, 
the Dirac Br\"{u}ckner Hartree--Fock theory reproduces nuclear saturation properties starting from two-body forces by solving
the Bethe--Salpeter equation, single-particle self-energy, and the Dyson equation \citep{bro90}.

\subsection{Composition inside neutron stars}

\begin{figure*}[ht]
\centering
\includegraphics[width=0.85\textwidth]{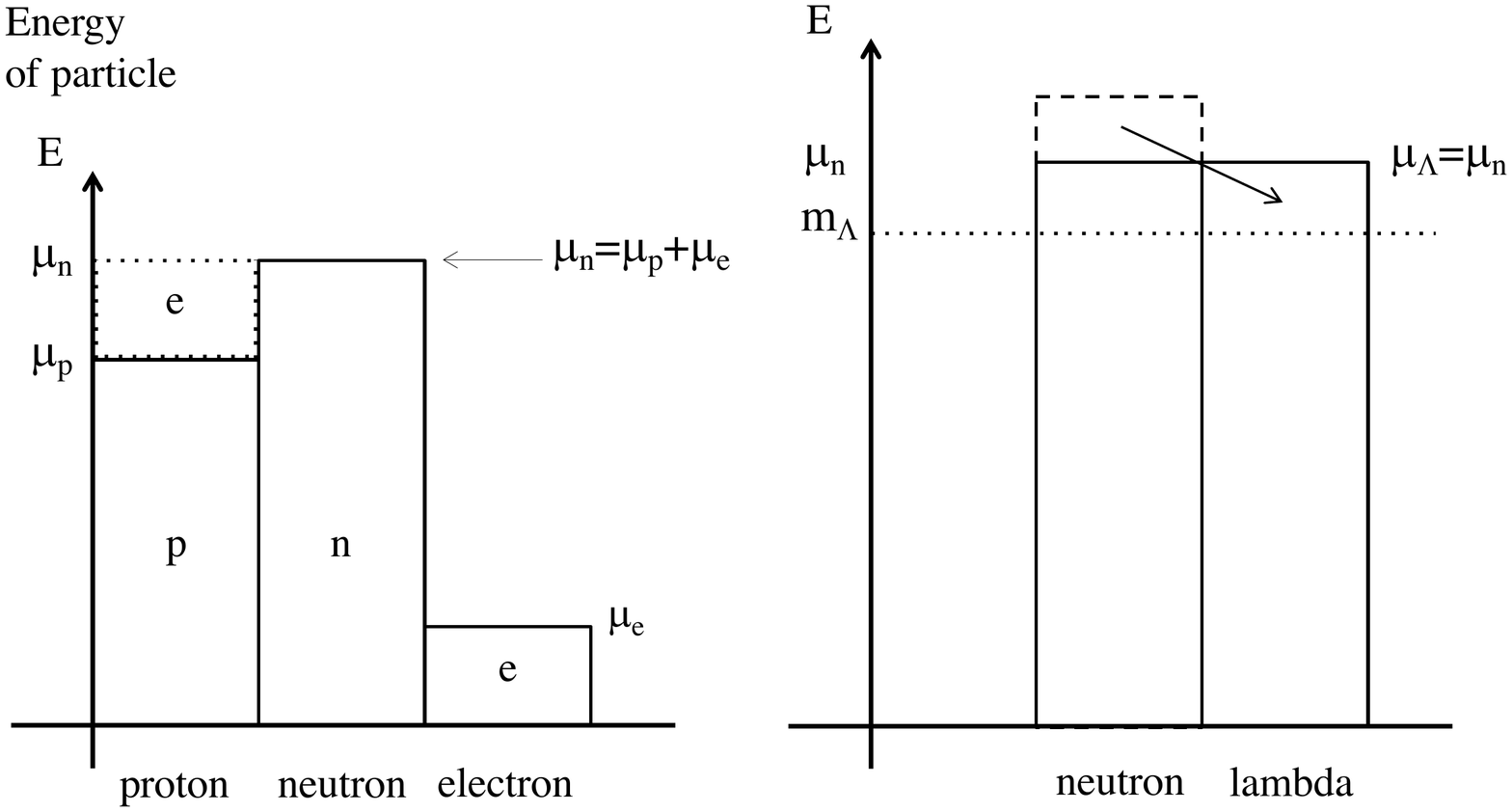}
\vspace{-1.8cm}
\caption{Schematic diagram of the occupied energy states of the degenerate particles in the neutron star matter.  Each particle fills the state up to the chemical potential (Fermi energy) under the beta equilibrium (left).  When the neutron chemical potential exceeds the hyperon mass, neutrons are converted to hyperons to reduce the total energy and fulfill the chemical equilibrium with hyperons (right)  \citep{KSBook2018}.  
\label{fig:NS_Fermi}}
\end{figure*}

In neutron star matter the charge neutrality and beta-equilibrium conditions introduce considerable asymmetry in the isospin density \citep{Shapiro_Book}.
The proton fraction is strongly correlated with the symmetry energy which characterizes the energy cost from the isospin asymmetry.
Neutron star matter at $n_0$ has a small proton fraction of $Y_p \sim 0.1$, which is obtained by minimizing the total energy density,
$\rho_{total}(n)= \rho (n,Y_p)+ \rho_e(n_e)$, 
with respect to the proton fraction, where $\rho_e(n_e)$ and $n_e = Y_p n$ are the electron energy density and number density.
At higher density muons can also contribute.  
The condition at the minimum corresponds to the relation of the chemical equilibrium among particles, $\mu_n=\mu_p+\mu_e$, which states the balance of the Fermi energy of neutrons versus those of protons and electrons, $n \leftrightarrow p+e^{-}$, as shown in the left panel of Fig. \ref{fig:NS_Fermi}.  
The set of these conditions is called as the beta equilibrium.

\begin{figure*}[ht]
\centering
\includegraphics[width=0.8\textwidth]{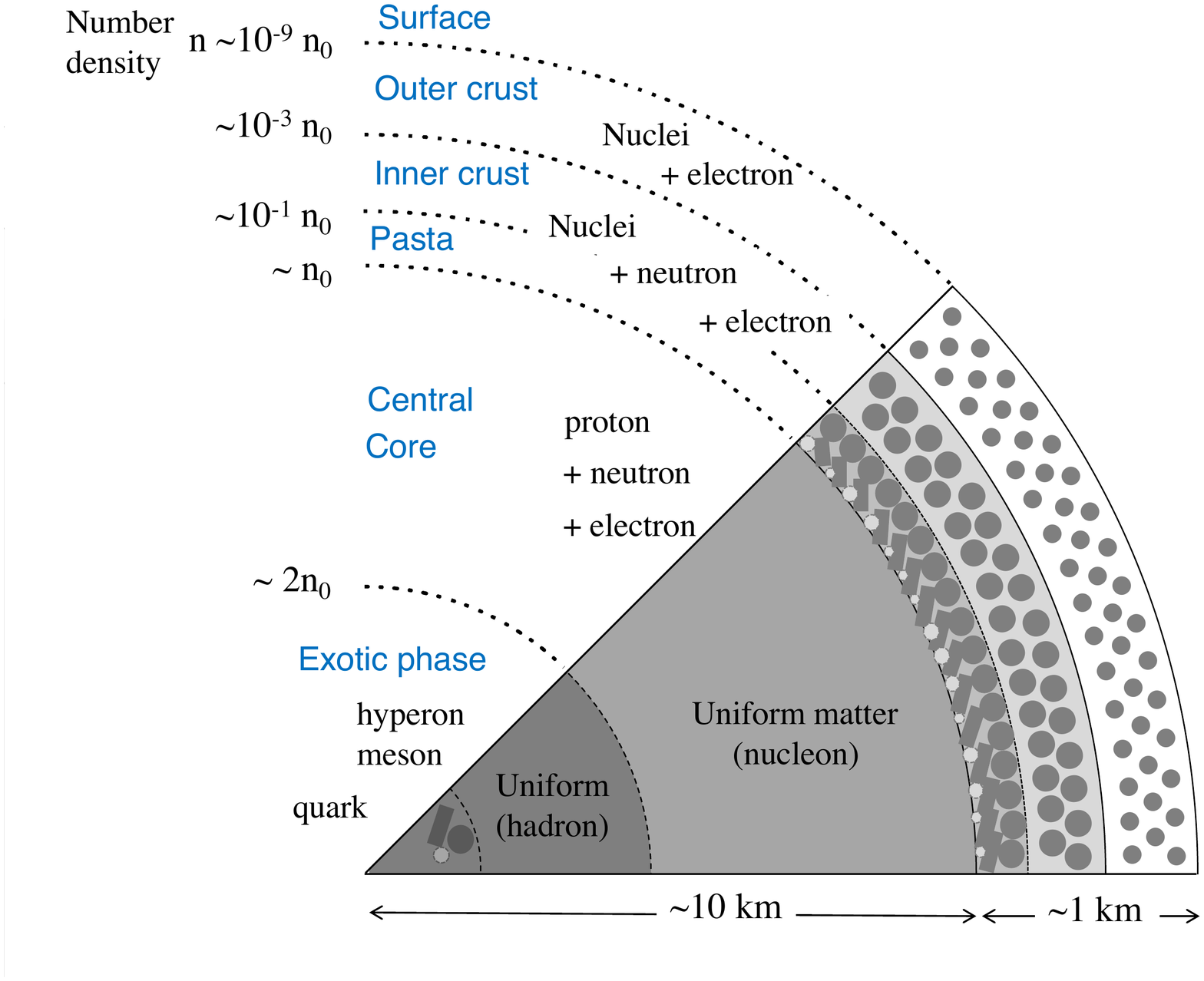}
\caption{Schematic diagram of the composition and the phase of matter in the interior of neutron star.  \citep{Heiselberg2002,KSBook2018}.  
\label{fig:NS_profile}}
\end{figure*}

\begin{figure*}[ht]
\centering
\includegraphics[width=0.38\textwidth]{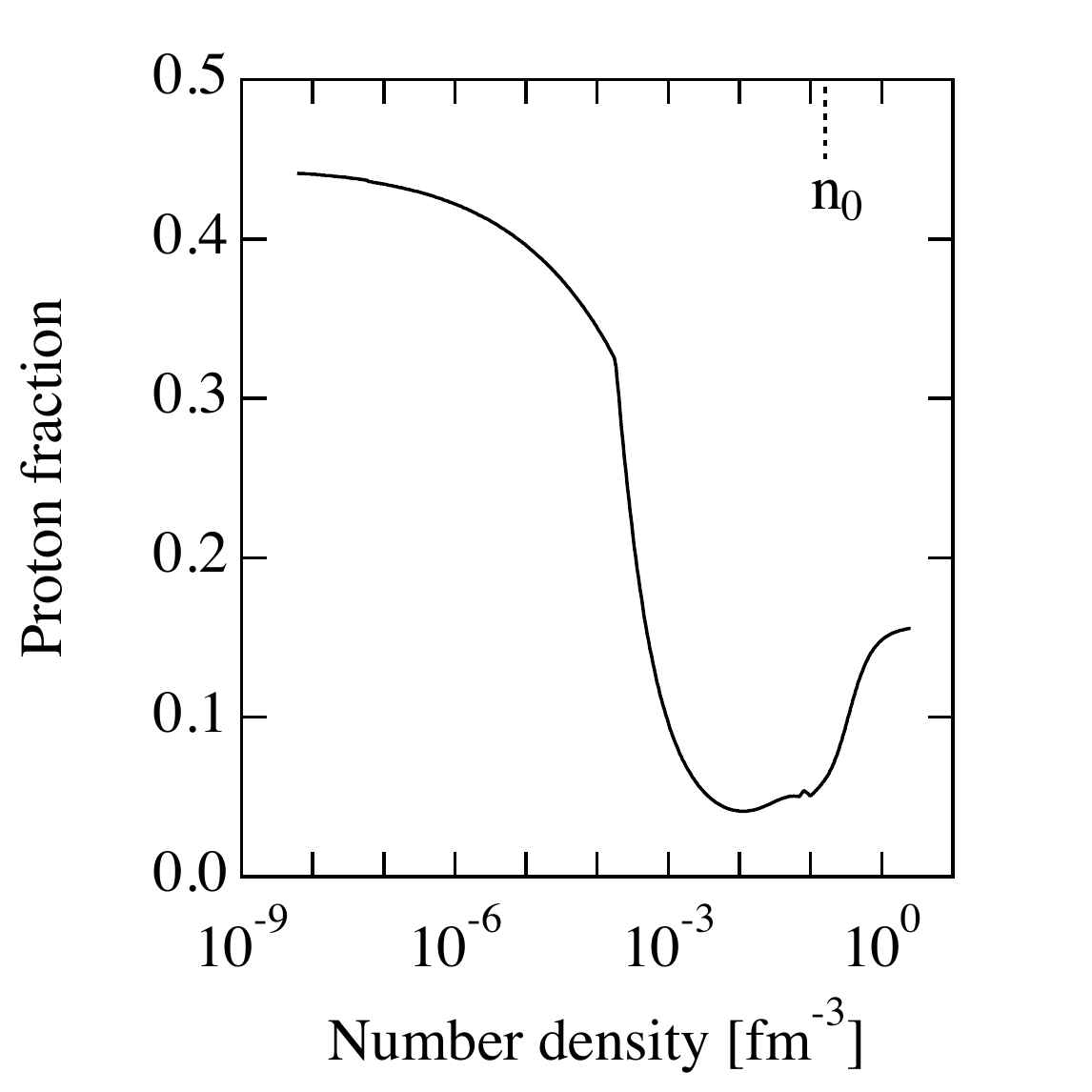}
\includegraphics[width=0.61\textwidth]{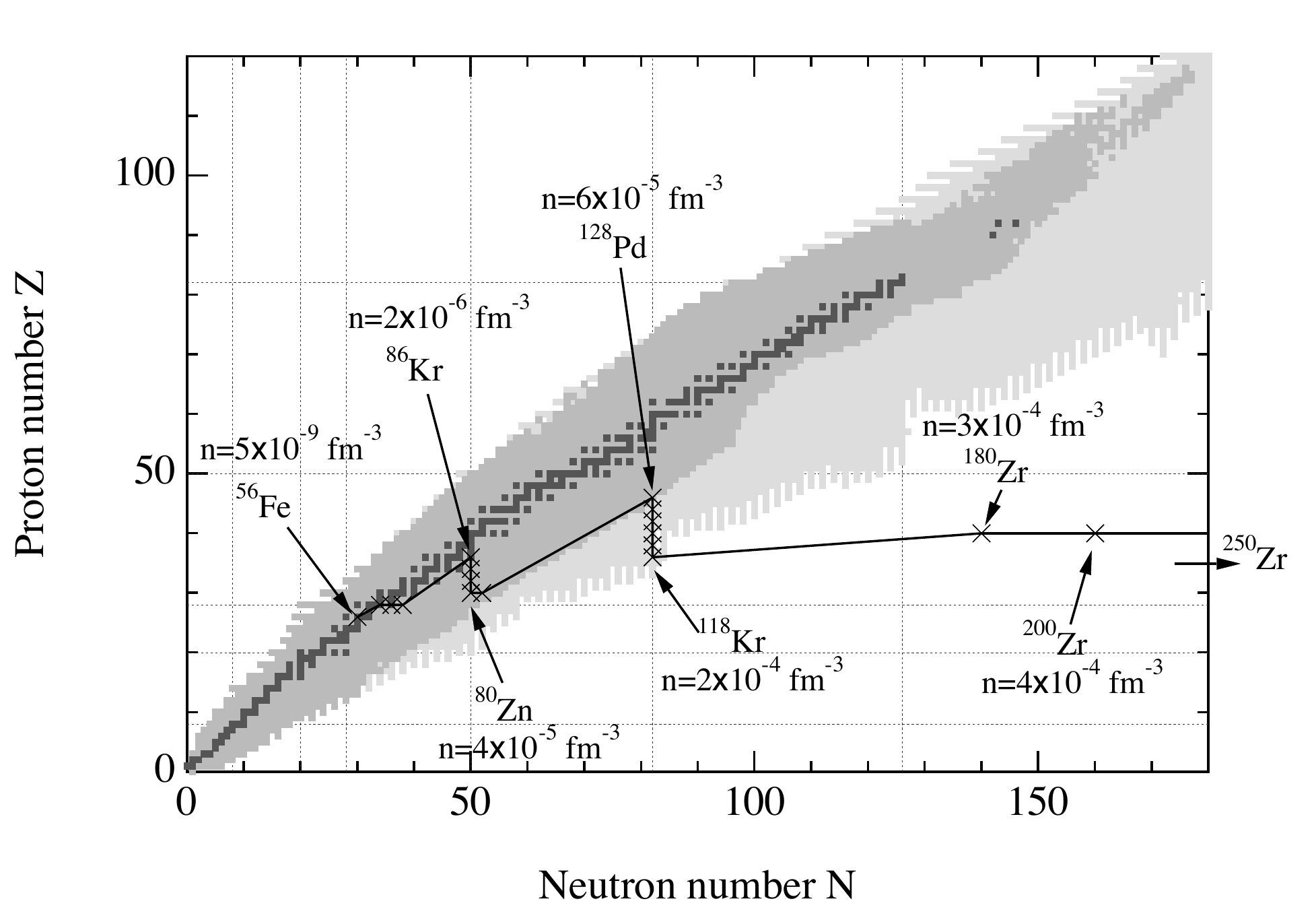}
\caption{Proton fraction of the neutron star matter as a function of number density (Courtesy of H. Shen) \citep{she20} (left) and species of nuclei in the neutron star matter in the nuclear chart (Courtesy of H. Koura)  \citep{Chamel2008,KSBook2018} (right).  
\label{fig:NS_composition}}
\end{figure*}

The composition and proton fraction change considerably from the surface to the core of a neutron star, 
where the density changes from $\sim 10^{-10}n_0$ to $\sim n_0$ as shown in Figs. \ref{fig:NS_profile} \citep{Heiselberg2002,KSBook2018} and \ref{fig:NS_composition} \citep{Chamel2008,she20}.  
Near the surface,  nucleons form a nucleus and electrons are localized around it.
The most stable nuclei are $^{56}$Fe and the corresponding proton fraction is $Y_p \simeq 0.46$.
This dilute regime begins to be modified beyond $10^{-8} n_0$, 
where it is energetically more favorable to reduce the number of electrons by the process $p+e^{-} \rightarrow n + \nu_e$.
Accordingly the proton fraction decreases and a number of neutron rich nuclei, as indicated in the right panel of Fig. \ref{fig:NS_composition}, appear .
This region is called the outer crust \citep{Chamel2008}.  
Around $10^{-3}n_0$, there are too many neutrons and they begin to drip out of nuclei.
The matter consists of nuclei, neutrons, and electrons in the region called the inner crust \citep{Chamel2008}. 
Further compression of nuclei to densities beyond $5\times 10^{-2}n_0$ merges them into the pasta structure, 
which has huge nuclei with various non-spherical shapes \citep{Oyamatsu1993}.  
Above the nuclear density, $\sim 0.5n_0$, the nuclei are dissolved into neutrons and protons and the matter becomes uniform.  
In this regime the cost associated with the symmetry energy dominates over the cost of having electrons, thus $Y_p$ grows as density does.

Deep inside the central core at $n \gtrsim 2n_0$, 
there may be exotic phases with hyperons or quarks.  
Hyperons contain strange quarks.
The appearance of these new particles is controlled by the condition of the chemical equilibrium.  
The chemical potential of neutrons, $\mu_n$, increases as the density goes up.  
When $\mu_n$ exceeds the mass of hyperons, $m_{\Lambda}$, for example, 
neutrons can be converted to hyperons because it reduces the Fermi energy of neutrons (right panel of Fig. \ref{fig:NS_Fermi}).  
Allowing the appearance of new particles at a given density usually softens equations of state, 
as it increases the energy density by $\Delta \rho \sim mn$ but the associated increase in pressure is much smaller, $\Delta P \sim n^{5/3}/m$.

\begin{figure*}[ht]
\centering
\includegraphics[width=0.45\textwidth]{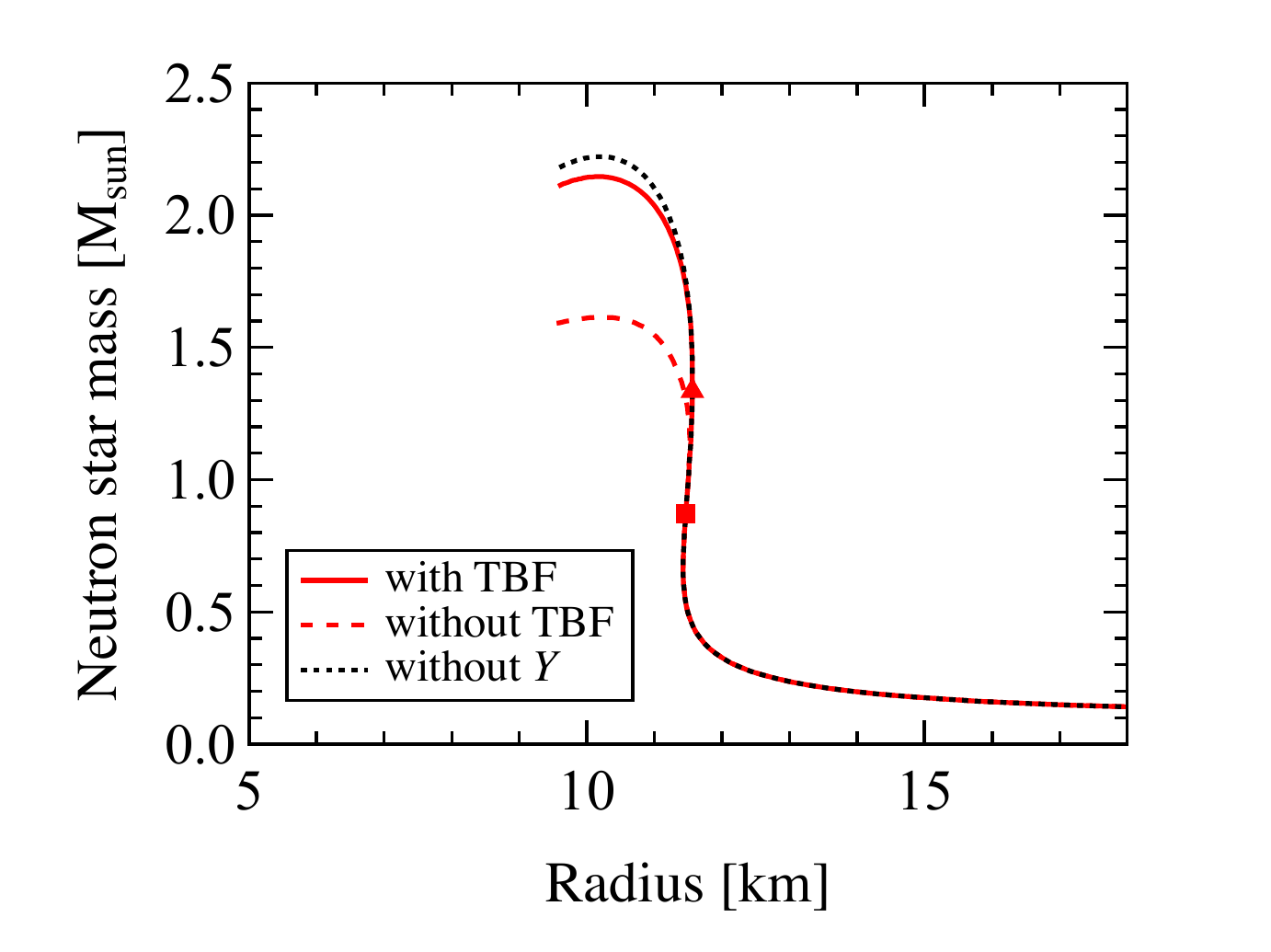}
\includegraphics[width=0.45\textwidth]{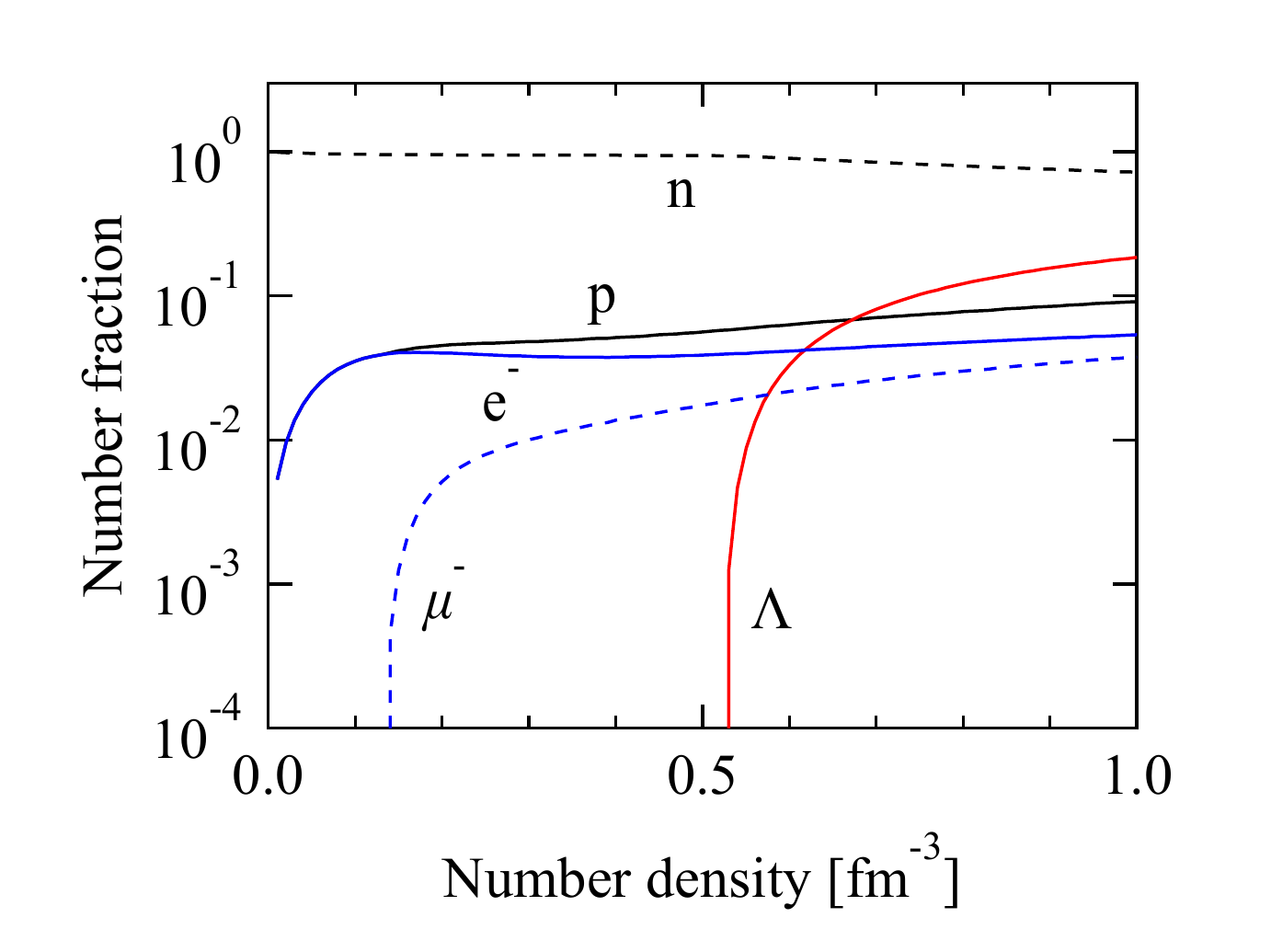}
\caption{(left) Gravitational mass of neutron stars constructed by the equation of state based on the cluster variational method \citep{tog16} 
are shown for a choice of repulsive $\Lambda\Lambda$ interaction as functions of the radius in the left panel.  
The models for hyperonic matter are shown by red solid and dashed lines with and without three body force for hyperons, respectively.  
The model without hyperons is plotted by black dot-dashed line.  
The onset of $\Lambda$ mixture is marked by symbols.  
Number fractions of the neutron star matter for a choice of repulsive $\Lambda\Lambda$ interaction with three body force are shown as functions of the density in the right panel.  
The solid red lines shows the number fraction of $\Lambda$.  (Courtesy of H. Togashi)
\label{fig:togashi_ns_hyperon}}
\end{figure*}

Shown in Fig. \ref{fig:togashi_ns_hyperon} are examples of $M$-$R$ curves for equations of state with and without hyperons,
based on equations of state calculated in variational methods \citep{tog16}.
The maximum mass of $\sim 2.2M_\odot$ for purely nucleonic equations of state drops to $\sim 1.6M_\odot$ after including hyperons together with reasonable $YN$ forces.
The small maximum mass incompatible with the $2M_\odot$ constraints suggests the necessity of some additional mechanisms.
One possibility is the existence of strong $YNN$ repulsions that increase the density at which hyperons appear, as shown in Fig. \ref{fig:togashi_ns_hyperon}.
Clearly it is important to examine the properties of $YN$ and $YNN$ interactions \citep{Vidana2018,Tolos2020}.
The experimental examination has difficulties since hyperons are unstable with respect to the weak decays and cannot be studied as in $NN$ scattering experiments.
One possible way to study $YN$ interactions is to inject hadrons with strangeness into nuclei, create hyper nuclei, and then study the spectroscopy \citep{Hiyama:2018lgs}. 
Another method, which has been developed in last ten years, is to measure $YN$ interactions in lattice QCD \citep{HALQCD:2018qyu,HALQCD:2019wsz}.
The $YNN$ forces have not been determined yet.

The composition of matter is important not only for $M$-$R$ relations, 
but also for non-mechanical aspects of neutron stars, such as cooling curves or chemical reactions \citep{Yakovlev2004,Page2004,Potekhin2015}.
The neutron stars cool down through the emission of neutrinos in dense matter, and its cooling rate strongly depends on the matter composition.
The fastest cooling mechanism in $npe\mu$ matter is the direct Urca processes 
in which $n \rightarrow p + e^{-} +\bar{\nu}_e$ and $p+e^{-} \rightarrow n + \nu_e$ processes produce neutrinos escaping from the neutron star cores \citep{Lattimer1991dU}.
The condition for these processes to occur is $Y_p \gtrsim 0.1$, as can be derived from the energy and momentum conservations.
If this condition is not met, the modified Urca processes, $n + n \rightarrow n + p + e +\bar{\nu}_e$ and $n + p + e \rightarrow n + n + \nu_e$, 
are the next candidates for cooling mechanisms.
The modified Urca is much slower than the direct one as the former requires two thermally excited nucleons, 
whose population are suppressed by the Boltzmann factor, to interact.
Due to this large difference in cooling time scale for the $Y_p \gtrsim 0.1$ and  $Y_p \lesssim 0.1$ cases, 
nuclear equations of state with the large symmetry energy can be differentiated from the others \citep{Page1992}.
The complication is that typical nuclear equations of state lead to $Y_p \gtrsim 0.1$ only at sufficiently high densities where hyperons or quarks may appear. 
They open new channels for the fast cooling.
Another important effect is the pairing gap which suppresses the abundance of thermally excited particles and thus neutrino production \citep{Chamel2008}.  
In what follows, the fast cooling requires sufficiently high density. 
The observed cooling curves seem to be consistent with the modified Urca, 
but it remains to be shown whether the core density reaches the density threshold for the direct Urca or not;
for now the cooling curves and neutron star masses are not measured simultaneously.

\section{\textit{Matter in Core-Collapse Supernovae}}

\subsection{Evolution of matter and neutrinos}
\begin{figure*}[ht]
\centering
\includegraphics[width=0.99\textwidth]{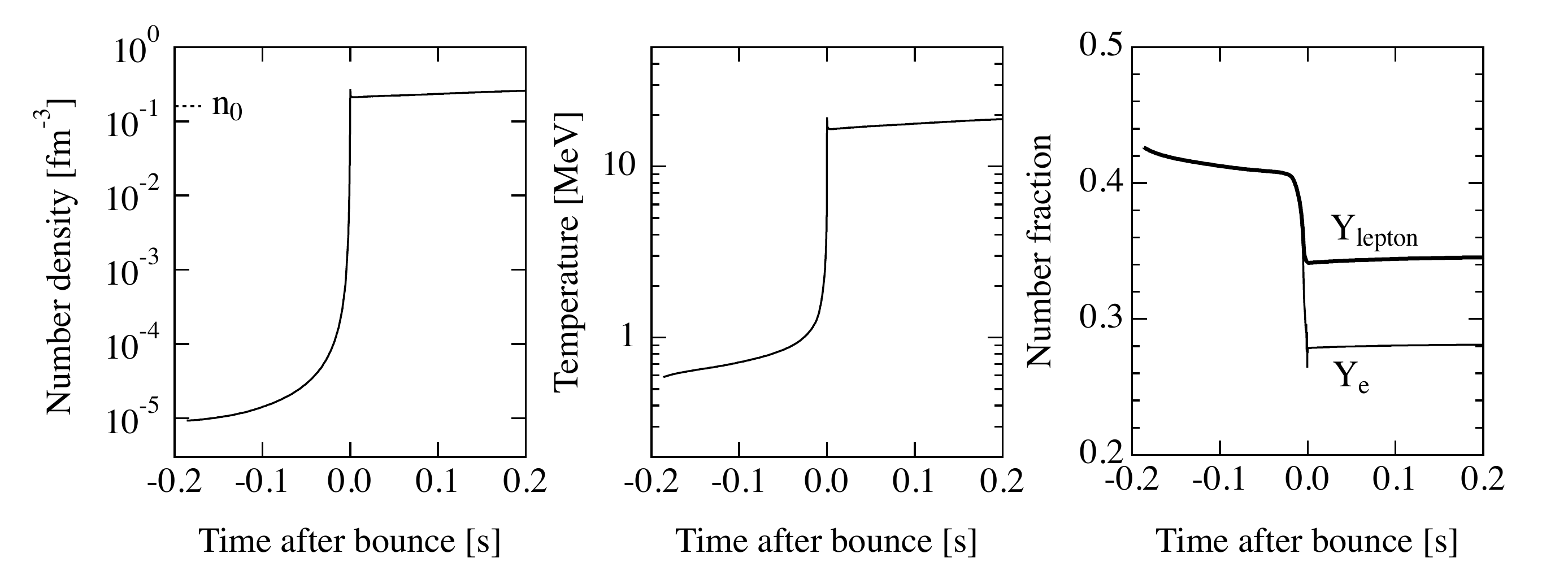}
\caption{Time evolution of the number density, temperature and number fractions of electrons and leptons during the core collapse and bounce are shown in the left, middle, and right panels, respectively.  
The quantities at the center are taken from the numerical simulation with a 11M$_{\odot}$ star.  
The time is measured from the timing at the core bounce.  
The lepton and electron fractions are shown by thick-solid and solid in the right panel, respectively \citep{fur17b,nak21}.
\label{fig:radhyd_profile_center}}
\end{figure*}

The properties of hot and dense matter in core-collapse supernovae drastically change in the dynamical situations of collapse, bounce and explosion \citep{Oertel:2016bki}.  
Typical conditions of the evolution of number density, temperature and electron fraction ($Y_e=Y_p$) at the center in the supernova core is shown in Fig. \ref{fig:radhyd_profile_center}.  
The density and temperature increase rapidly in the collapse and reach high values at the core bounce.  
The electron fraction $Y_e$ decreases and remains $\sim0.3$.  
It is useful to understand the time scale of changes of conditions and to examine the response of matter and neutrinos in the evolving environment.  

The matter composed of nucleons, nuclei, electrons, positrons and photons is treated as a fluid component.  
The dynamical time scale, $T_{dyn}$, in fluid motion of free-fall under the gravitational force can be estimated to be $T_{dyn} \sim 5\times10^{-3}\,{\rm s} \, (n/10^{-4}\, {\rm fm}^{-3})^{-1/2}$ \citep{Shapiro_Book}.  
It takes $\sim0.2-0.3$ s from the start of the gravitational collapse to the core bounce.  
This is much longer time scale than in strong and electro-magnetic interactions.  
Therefore, the nuclear and electro-magnetic processes can proceed fast enough to achieve the thermal and chemical equilibrium.  
Equations of state at a certain density, temperature, and electron fraction are necessary to perform numerical simulations.  

Neutrinos are treated separately by solving the equations of neutrino transfer, which describes the propagation and reaction of neutrinos in matter. 
The time scale of weak reactions can be comparable or even longer than the dynamical time scale.  
For example, the time scale of the electron capture on free proton can be estimated to be 
$T_{e-cap} \sim 3\times10^{2}~{\rm s}\, (n/10^{-6}~{\rm fm}^{-3})^{-5/3}(Y_e/0.46)^{-2/3}(X_p/10^{-4})^{-1}$ \citep{Shapiro_Book,suz94} 
where $X_p$ is the mass fraction of free protons and the degenerate relativistic electron gas is assumed.  
It is not possible to assume the equilibrium for weak reactions, being different from the case of cold neutron stars.  
It is mandatory to follow the time evolution of the electron fraction of matter by tracking the modifications of composition through weak reactions.  
This requires detailed descriptions of neutrino reactions in matter as a source of changes.  
It is also necessary to provide the weak reaction rates, which largely depend on the target particles and environment of the matter.

\subsection{Nuclear statistical equilibrium}
 
 At temperature of a few MeV achieved in supernova matter, fusing $N$ neutrons and $Z$ protons into a nuclide $(N,Z)$ is balanced
 by the photo-breakup reaction of the nuclide $(N,Z)$ into $N$ free neutrons and $Z$ free protons.
 In this nuclear statistical equilibrium, the compositions of nuclear matter (number densities of  nucleons and all nuclei) 
 are determined by minimizing the free energy of a model \citep{Mazurek1979,hem10}.
The free energy density of nuclear matter consists of contributions from nucleons and  various nuclei as
\begin{equation}
f=f_p+f_n+ \sum_{N,Z} f_N(N,Z),
\end{equation}
where  $f_p$ and $f_n$ are free energy densities of protons and neutrons, respectively.
The free energy density of nuclei, $f_N(N,Z)$, is expressed as 
\begin{equation}
f_N(N,Z)=n_N(N,Z)  \{M(N,Z) + F_t(N,Z) \}
\end{equation}
where $n_N(N, Z)$  represents the nuclear number density, $M(N, Z)$ represents nuclear mass
and $F_t(N,Z)$ represents the free energy of translational motion. 

The number densities of nuclei under the nuclear statistical equilibrium for the given $\rho$, $T$, and $Y_p$ 
are obtained by minimizing the model free energy with respect to the variational parameters.
In the minimization the following mass and charge conservations are imposed,

\begin{equation}
 n_p+n_n+\sum_{N,Z}{(N+Z) n_N }=n  ,  \label{eq_const1}
\end{equation}
\begin{equation}
 n_p+\sum_{N,Z}{Z n_N}=Y_p n,  \label{eq_const2}
\end{equation}
where  $n_p$ and $n_n$ are the number densities of free protons and neutrons, respectively. 
 By introducing Lagrange multipliers $\alpha$ and $\beta$ for the above constraints,
the free energy minimization as a function of $n_p$ and $n_n$ leads to ($i=p,n$)
\begin{eqnarray}
 \frac{\partial}{\partial {n_i}} 
 	\bigg[f - \alpha \bigg( n_p+n_n+\sum_{N,Z}{(N+Z) n_N } - n  \bigg) - \beta \bigg(n_p+\sum_{N,Z}{Z n_N }-Y_p n \bigg) \bigg] =0,  
\end{eqnarray}
Thus, the Lagrange multipliers are expressed with the chemical potentials of protons  and neutrons, $\mu_p$ and $\mu_n$, as 
\begin{eqnarray}
 \alpha & = &\frac{\partial f}{\partial n_n}= \mu_n, \\
 \beta & = & \frac{\partial f}{\partial n_p} -  \frac{\partial f}{\partial n_n}= \mu_p -\mu_n.
\end{eqnarray}
The differentiation with respect to $n_N$, leads to  the chemical 
potentials of nuclei, $\mu(N, Z)=\partial f/\partial n_N$, as 
\begin{equation}
\label{eq:nse}
 \mu(N,Z)=N \mu_n + Z \mu_p.
\end{equation}
Here, nucleons and nuclei are treated as the ideal Boltzmann gases with their constant masses as
\begin{eqnarray}
f(N, Z) &=& n_N(N, Z)  \left[  M(N, Z)  +T \ln \left\{ \frac{n_N(N,Z)}{g(N, Z) n_{Q}(N,Z)} \right\}  - T \right]  \label{eq_bg} , \\
 n_{Q}(N,Z) & = &  \left(\frac{M(N,Z) k_B T}{2 \pi} \right)^{3/2} . 
\end{eqnarray}
The partial differential of Eq.~\ref{eq_bg} with respect to $n_N(N, Z)$ and Eq.~(\ref{eq:nse})
lead to  $n_N(N, Z)$ as follows:
\begin{eqnarray}
n_N(N, Z) = g(N, Z)\, n_{Q}(N,Z) \exp \left( \frac{Z \mu_p+ N \mu_n - M(N, Z)}{T} \right), \label{eq:nin} 
\end{eqnarray}
where   $g(N, Z)$ is degeneracy factor of the nucleus. 
Calculations  of  $M(N,Z)$ depend on equations of state \citep{fur20b}.

\subsection{Collapse of the Fe core and weak interactions}

Environment in supernovae evolves from a gravitationally collapsing massive star to the birth of a neutron star \citep{bet90,jan12a,oer17}.  
The gravitational collapse starts from the central Fe core in the final stage of a massive star with the mass greater than $\sim 10M_{\odot}$.  
The central number density and temperature is typically $\sim10^{-5}$~fm$^{-3}$ and $1$~MeV in the Fe core, 
which has a mass of $\sim 1.5M_{\odot}$ and radius of $\sim10^4$ km, supported by pressure of electrons.  
The main composition is nuclei around $^{56}$Fe which has a proton fraction, $Y_p=Z/A=26/56=0.46$, 
as the final product of nuclear fusion reactions to the most bound nuclei.  
Above the temperature $\sim1$~MeV, the nuclear reactions among nucleons and nuclei proceed fast enough fast to maintain the 
nuclear statistical equilibrium.  

\begin{figure*}[ht]
\centering
\includegraphics[width=0.9 \textwidth]{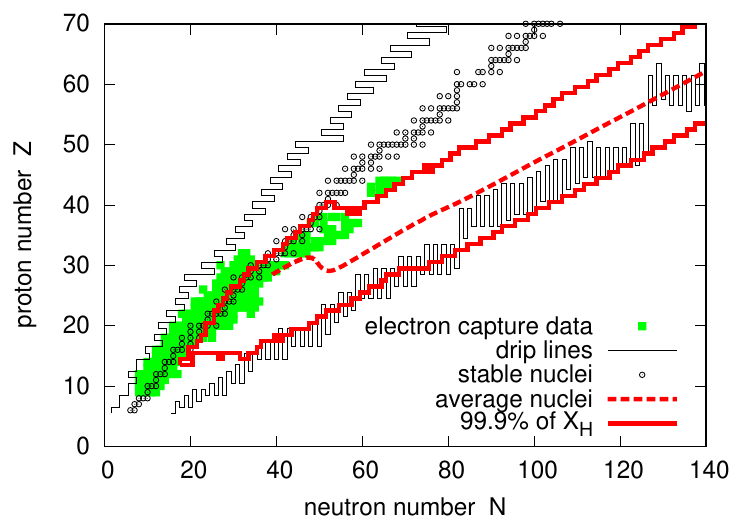}
\caption{
The nuclear species for which the data of  electron capture rate is available (green squares). Black circles represent stable nuclei and solid thin black lines display the neutron and proton drip lines. 
A dashed red line indicates the trajectory of average neutron and proton numbers of heavy nuclei.
Nuclei inside the solid thick red line appear in the center of collapsing cores and account for the top 99.9\%  of total mass fraction of heavy nuclei at any time step  \citep{fur17c}. 
\label{fig:nse_core_collapse}}
\end{figure*}

The matter becomes neutron-rich in the collapse of the central core (Fig. \ref{fig:radhyd_profile_center} right). 
The electron captures proceed due to the high Fermi energy of electrons and the electron fraction decreases.  
The distribution of nuclei is shifted to the neutron-rich side, being away from the stability line. 
Fig.~\ref{fig:nse_core_collapse} displays the nuclei that are abundant during the collapse.
The nuclear reactions during the collapse depend on the properties of neutron-rich nuclei and the associated response to electron captures.  

At the initial stage, the neutrinos produced by the electron captures freely escape from the central core.
They carry away the lepton number which is originally carried by electrons. 
As the electron captures proceed and neutrinos fly away,
the electron fraction in the core decreases.

Further collapse of the central core, however, leads to higher densities of matter and the neutrino trapping due to the frequent scattering of neutrinos on nuclei.  
The neutrino transport eventually proceeds with the diffusion process.  
The time scale for the diffusion, $T_{diff}$, 
in the central core is estimated to be $T_{diff} \sim 4\times10^{-2}~{\rm s} \,(n/10^{-4}~{\rm fm}^{-3})(A/56)$ based on the neutrino mean free path \citep{Shapiro_Book,suz94}.  
It becomes longer than the dynamical time scale, $T_{dyn} \sim 5\times10^{-3}$~s at the density,  $\rho=10^{-4}$~fm$^{-3}$ \citep{Shapiro_Book,suz94}.
Therefore, the neutrinos cannot escape anymore and are trapped inside the central core beyond this density.  

In matter compressed together with neutrinos, the trapped neutrinos can be treated as degenerate leptons.  
This state of matter is often called as the supernova matter, which is in thermal and chemical equilibrium.  
The supernova matter is parameterized by the lepton fraction, $Y_L$, instead of the electron fraction, $Y_e$.  
After the neutrino trapping, the lepton fraction remains constant during the gravitational collapse and after the core bounce (Fig. \ref{fig:radhyd_profile_center} right).
The lepton fraction at this stage is important to determine the size of the bounce core to launch the shock wave.  

Since there is no escape of particles to carry energy, the matter is compressed under the adiabatic condition.  The entropy per baryon remains constant after the neutrino trapping during the gravitational collapse.  The rise of the temperature is determined by the properties of the matter along the adiabatic curve.  
The temperature increases moderately even though the density increases dramatically (Fig. \ref{fig:radhyd_profile_center} middle). 

\subsection{Core bounce toward the explosion}
\begin{figure*}[ht]
\centering
\includegraphics[width=0.9\textwidth]{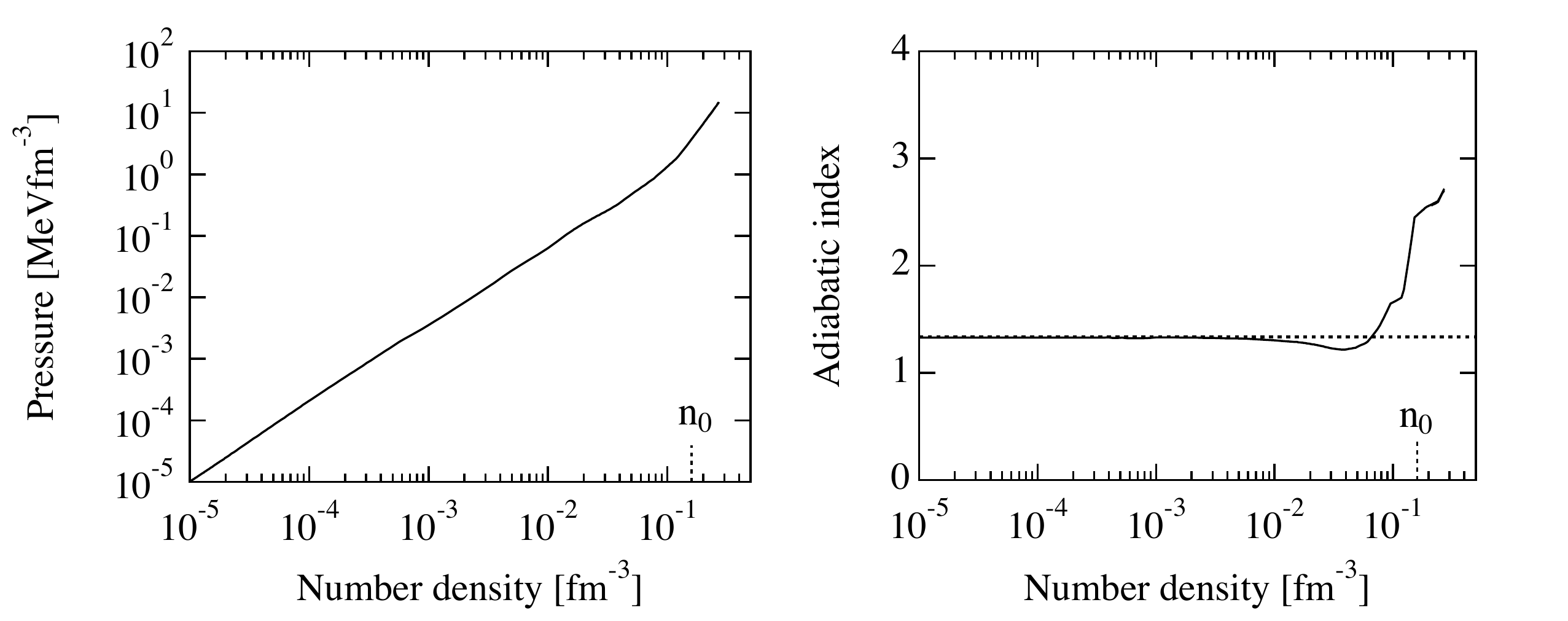}
\caption{Pressure and adiabatic index of the matter are shown in the left and right panels, respectively, as a function of the number density.  
The slope of pressure steepens above $\sim n_0$ and the adiabatic index increases suddenly.  \citep{fur17b,nak21}
\label{fig:radhyd_eos_gamma}}
\end{figure*}

\begin{figure*}[ht]
\centering
\includegraphics[width=0.9\textwidth]{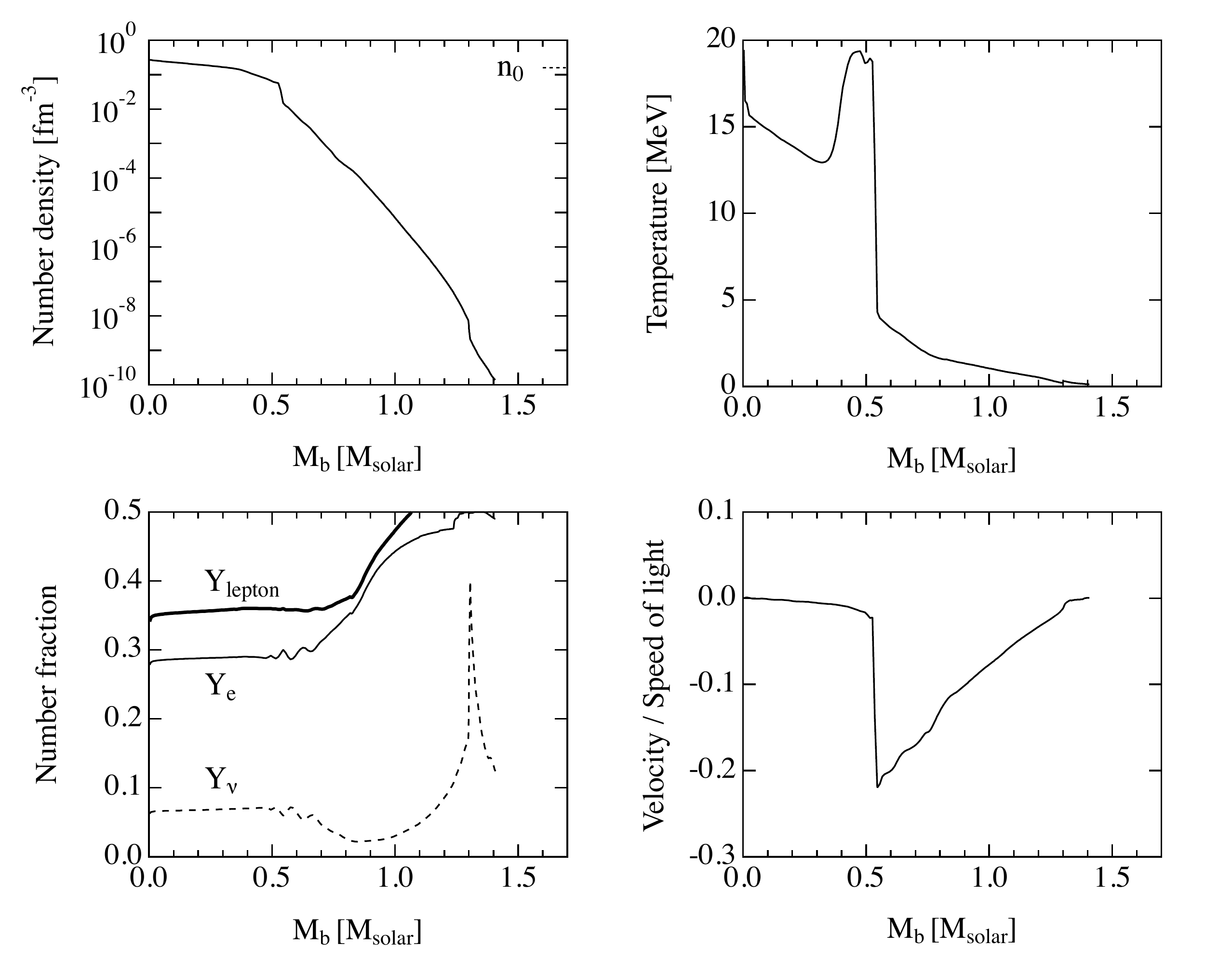}
\caption{ Number density, temperature, number fractions of electrons and leptons, and velocity at the core bounce are shown as a function of the radial coordinate measured in mass.  
Profiles are taken from the numerical simulation of gravitational collapse of 11$M_{\odot}$ star.  \citep{fur17b,nak21}
\label{fig:radhyd_profile_tpb0}}
\end{figure*}

The gravitational collapse with the neutrino trapping 
halts suddenly just above the nuclear density, $n_0$.  
The pressure abruptly increases above the nuclear density and the equation of state becomes stiff as shown in Fig. \ref{fig:radhyd_eos_gamma}.  
This is mainly due to the repulsive contribution of nuclear forces.  
The adiabatic index, which corresponds to the slope of pressure versus density, becomes large $\sim2$ above the value $\sim$4/3 
for lepton gas during the collapse.   
This stiffening brings to a sudden stop of the compression of matter at the center and leads to bouncing back of the inner core.  
The shock wave is launched at the core bounce and starts propagating toward the outside as shown in Fig. \ref{fig:radhyd_profile_tpb0}.  
The central density is slightly above the nuclear density and the temperature is around $10-20$ MeV.  
The electron and lepton fractions are $\sim$0.29 and $\sim$0.35 due to the neutrino trapping with neutrino fraction $\sim$0.06 in this example.  
Note that the matter is not so neutron-rich having the proton fraction $\sim$0.29.  

This is the starting point of the explosion although there are a number of obstacles to overcome.  
The initial position of the shock wave is located in the middle of the Fe core.  
The shock wave must reach the surface of the Fe core by propagating against the free-fall of matter of the outer part.  
It slows down also due to the loss of energy by the dissociation of Fe-group nuclei.  
The shock wave eventually stalls during the propagation inside the Fe core.  
This is the second stage of the explosion typically seen in many simulations, especially under the spherical symmetry, which leads to the failure of explosion.  

\begin{figure*}[ht]
\centering
\includegraphics[width=0.9\textwidth]{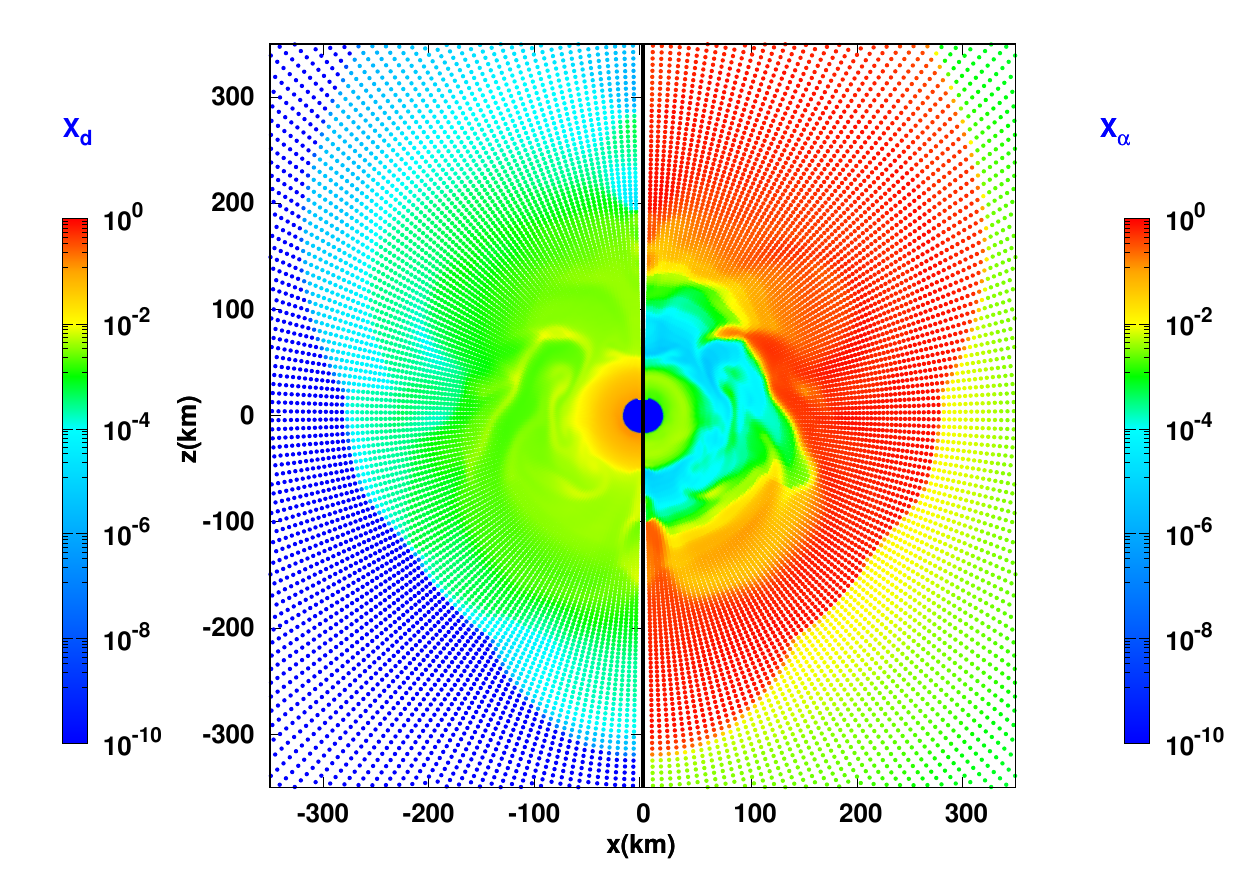}
\caption{Distribution of  deuteron mass fractions  (left) and $\alpha$ particle mass fractions  (right)  at  200 ms  after core bounce in a 2D simulation \citep{nag19c}.  
\label{fig:nse_shock}}
\end{figure*}

In the current understanding of the explosion mechanism, it is considered that the neutrino heating mechanism 
assists the revival of the stalled shock wave \citep{bet85,bet90}.  
The trapped neutrinos in the central part are gradually emitted to the outside and a part of outgoing neutrinos are absorbed 
by the material in the heating region just behind the stalled shock wave.  
The absorption of neutrinos heats the matter, increasing the internal energy and pressure to push the shock wave outwardly.  
The matter below the shock wave is composed of nucleons and light nuclei such as deuterons and $\alpha$ particles.
 Fig.~\ref{fig:nse_shock} displays the mass fractions of the nuclei.
 The deuterons are abundant at $R\sim 10$--$50$~km above the surface of the proto-neutron star,
 while $\alpha$ particles are available at $R \sim 100$--$300$~km around and inside the shock wave. 
 The nuclear matter that consists of nucleons and light nuclei also appear in heavy ion collisions
 and  equations of state at low densities, $0.01 n_0 \lesssim n \lesssim 0.2 n_0$, and may be experimentally constrained
 \citep{qin12,Hempel2015}.
 
The success of explosion requires an additional factor to raise the efficiency of the neutrino heating mechanism \citep{jan12b,bur13,kot13,jan16,jan17b}.  
In many of recent multi-dimensional simulations, the hydrodynamical instabilities assist the revival of the shock wave by giving an enough time for the neutrino heating.  The convective motion, for example, brings the matter hovering in the heated region behind the shock wave to absorb enough neutrinos.  
The combination of the neutrino heating and the hydrodynamical instability is believed to be the major scenario of the explosion mechanism 
although the outcome of the (non-)explosion widely depends on the numerical simulations and microphysics.  
  
\subsection{Influence of nuclear physics on supernovae}

Nuclear physics plays an important role to understand the initial stage of the shock wave during the collapse and bounce.
The initial shock energy at the core bounce is estimated from the gravitational binding energy of the bounce core to be $E^{ini}_{shock}\sim GM^2_{bounce}/R_{bounce}$ 
where $M_{bounce}$ and $R_{bounce}$ is the mass and radius of the bounce core.  
The mass of the bounce core is determined by the Chandrasekhar mass, $M_{Ch}=1.5(Y_L/0.5)^2 M_{\odot}$, 
supported by the pressure of leptons \citep{Shapiro_Book,suz94}.  
The lepton fraction, $Y_L$, is determined by the amount of electron captures and neutrino trapping during the collapse.  
The radius of the bounce core is determined by equations of state above the saturation density.  
For successful supernovae explosions, high compression of a matter and the associated quick increase in pressure is favored.
Equations of state which are soft at low density meet such condition,
but too soft ones are incompatible with the known properties of nuclei and neutron stars \citep{bar85,tak88}.
There are also intriguing possibilities that a strong quark-hadron first order phase transition triggers 
the second bounce to launch the shock wave for the successful explosion \citep{fis11}.

After the launch of shock waves, the nuclear physics is influential to the propagation of shock wave.  
One of the key factors is the mass of the bounce core; for a larger mass,
the shock wave propagation is less disturbed by the energy loss associated with dissociation of the Fe nuclei \citep{jan12b}. 
Another important physics is the neutrino energy at the emission region.
Efficient heating of the shock waves favors energetic neutrino fluxes, which can be produced by high compression and high temperature 
as achieved by soft equations of state.
Here the composition of matter is also very important, as the neutrino reactions strongly depend on target particles.  
Modeling of the nuclear statistical equilibrium and nuclear weak interactions  
has a significant impact on proto-neutron star masses and shock wave evolution to the same degree as the stiffness of the equation of state \cite{nag19a}.

\subsection{Birth of proto-neutron star}

\begin{figure*}[ht]
\centering
\includegraphics[width=0.99\textwidth]{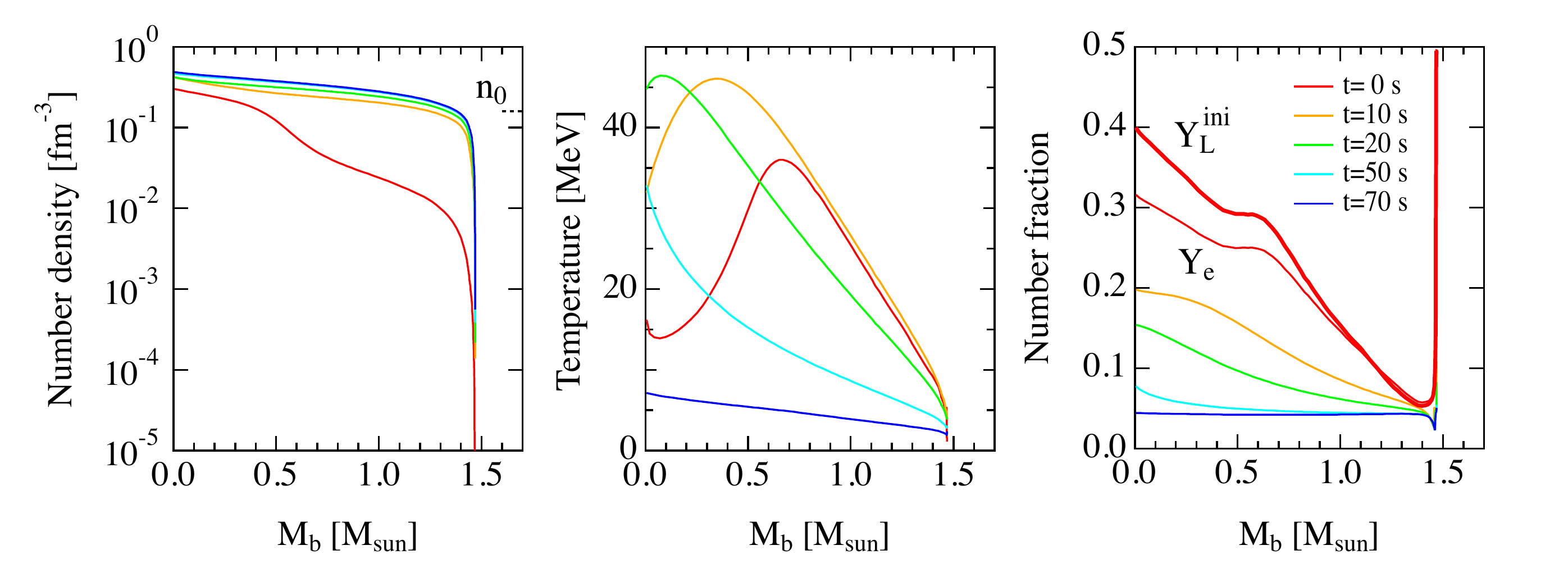}
\caption{Profiles of number density, temperature and number fractions of electrons at 0, 10, 20, 50, and 70 s are shown as a function of the radial coordinate measured in mass. 
The lepton fraction at 0 s is also shown.  
The time is measured from the start of the simulation, which corresponds to the timing $\sim0.3$~s after the bounce.  \citep{fur17b,nak21}
\label{fig:radhyd_pnsc_tx}}
\end{figure*}

After the launch of shock waves, the central part of the core settles down to the quasi-hydrostatic configuration and the compact object is born at the center $\sim$0.3 s after the core bounce.  
It is called proto-neutron star which contains a plenty of neutrinos and anti-neutrinos of all three flavors \citep{bur86,suz94,Pons99}.  
Those neutrinos are produced by the weak reactions including the thermal processes such as the pair creation.  
It is also hot ($T \gtrsim 10$~MeV) and less neutron-rich ($Y_e \lesssim 0.3$) with a large radius $\sim50$~km as compared to cold neutron stars as shown in the red lines in Fig. \ref{fig:radhyd_pnsc_tx}.  
The lepton fraction including neutrinos is determined during the collapse.

Proto-neutron stars cool down by emission of neutrinos.  
The temperature and electron fraction decrease since the neutrinos carry away the internal energy and lepton number as shown in Fig. \ref{fig:radhyd_pnsc_tx}.  
The evolution proceeds gradually over the time scale of $\sim20$~s through the diffusion of neutrinos in the matter.  
The proto-neutron star becomes compact with radii $\lesssim15$~km and the density becomes high.  
The proton fraction decreases to a small value, $Y_p\lesssim0.1$, which is determined by the beta equilibrium without neutrinos.  
The proto-neutron star turns into a cold neutron star over the time scale of minutes.

The supernova neutrinos contain various information to probe inside the supernova core \citep{Burrows1988,suz94,jan17a,BMueller2019}.  
The total energy of neutrinos can be used to derive the binding energy of the neutron star ($\sim10^{53}$~erg).  
The average energy roughly reflects the temperature at the emission region.  
The time duration of neutrino emission is related with the diffusion time scale determined by the density.  
For example, soft equations of state lead to high energies of neutrinos and a long duration due to high density and temperature.  
Note that the neutrino emission is closely related with the explosion mechanism.  
The dynamics due to non-spherical motion of matter such as the convection and/or rotation leads to rapid variations of neutrino emission in time and, therefore, may be probed by neutrinos \citep{BMueller2019}.  
Observation of supernova neutrinos may provide the information on the properties of neutrinos through the phenomena of neutrino oscillations \citep{Kotake2006,Duan2009,Mirizzi2016}.  
The neutrinos may affect also the explosive nucleosynthesis, which takes place in the outer layers, by changing the composition \citep{Woosley1990}.  

The supernova explosion is a target of the observation of gravitational waves, which brings the information of multi-dimensional dynamics \citep{Kotake2006,Ott2009,kot13}.  The time variation of the quadrupole moment of the matter distribution is a source of the variation of space-time metric in the Einstein equation of general relativity.  The proto-neutron star is excited by fluid motions to have oscillations with the eigen frequencies of the gravitational wave \citep{Ferrari2003,Sotani2016}.  
Simultaneous detection of the neutrinos and gravitational waves from the nearby supernova will help to reveal the explosion dynamics in the era of the multi-messanger astronomy \citep{yok14}.  




\subsection{Formation of black hole}
\begin{figure*}[ht]
\centering
\includegraphics[width=0.99\textwidth]{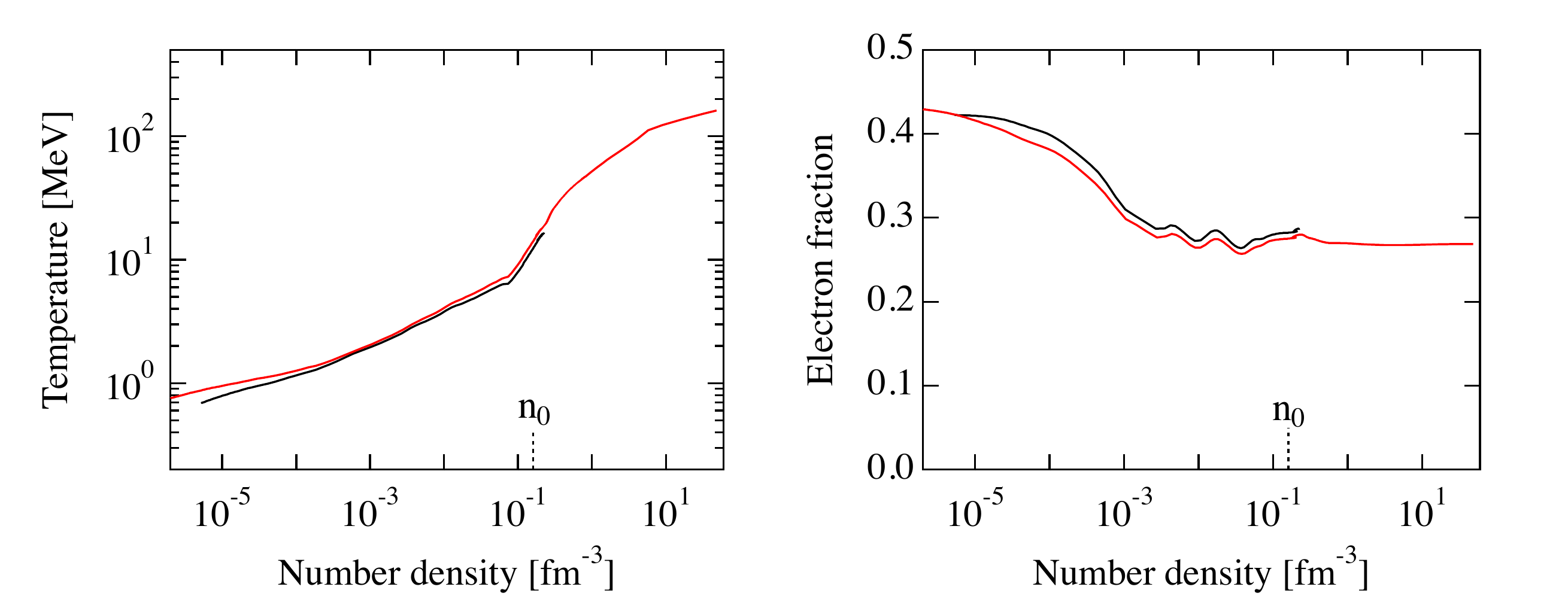}
\caption{Temperature and electron fraction distributions as a function of number density, around a core in the formation 
of a proto-neutron star (black lines) and a black hole (red lines).  
The data are taken from the numerical simulations of the collapse of 15$M_{\odot}$ and 40$M_{\odot}$ stars, 
which correspond to proto-neutron star and black hole formations \citep{sum19,she20,2017JPhCS.861a2028S}.  
\label{fig:radhyd_center_bh}}
\end{figure*}

The fate of the massive stars depends on the properties of the Fe core such as the compactness of density profile \citep{Heger2003,ocon11}.  
If the shock waves fail to revive, the accretion of matter continues and the mass of the proto-neutron star keeps growing.
When it goes beyond the maximum mass supported by the supernova matter, the dynamical collapse occurs and a black hole is formed.  
A massive star fades away as a failed supernova \citep{Kochanek2008,Adams2017}.  
In Fig. \ref{fig:radhyd_center_bh}, examples of the condition of matter in the evolution of supernova cores are shown.  
In the case of black hole formation, the density and temperature become very high 
and can be even beyond 100 MeV and $10$~fm$^{-3}$ due to the mass increase.  
The condition is more extreme than the case of proto-neutron star where the density stops growing at some point.  
This may open up possibilities of the appearance of the exotic particles such as hyperons and quarks.  
Note that the electron fraction is not so low (Fig. \ref{fig:radhyd_center_bh} right) due to the neutrino trapping, 
which tends to suppresses the appearance of exotic particles at the birth of proto-neutron star.  
Nevertheless, the evolution of density and temperature in the black hole formation can be extreme enough to yield hyperons and quarks.  

Since the energy release of the accreting matter is efficient and the temperature becomes high, 
the luminosity and average energy of neutrinos increase rapidly toward the black hole formation.  The typical duration of neutrino emission is $\sim1$~s under the spherical symmetry.  
These features may be distinguished from the ordinary supernova neutrinos.  
They can be used to probe equations of state at extreme conditions \citep{sum06}.  



\section{\textit{Matter in Merger of neutron stars}}

\begin{figure*}[ht]
\vspace{-1.cm}
\centering
\includegraphics[width=0.90\textwidth]{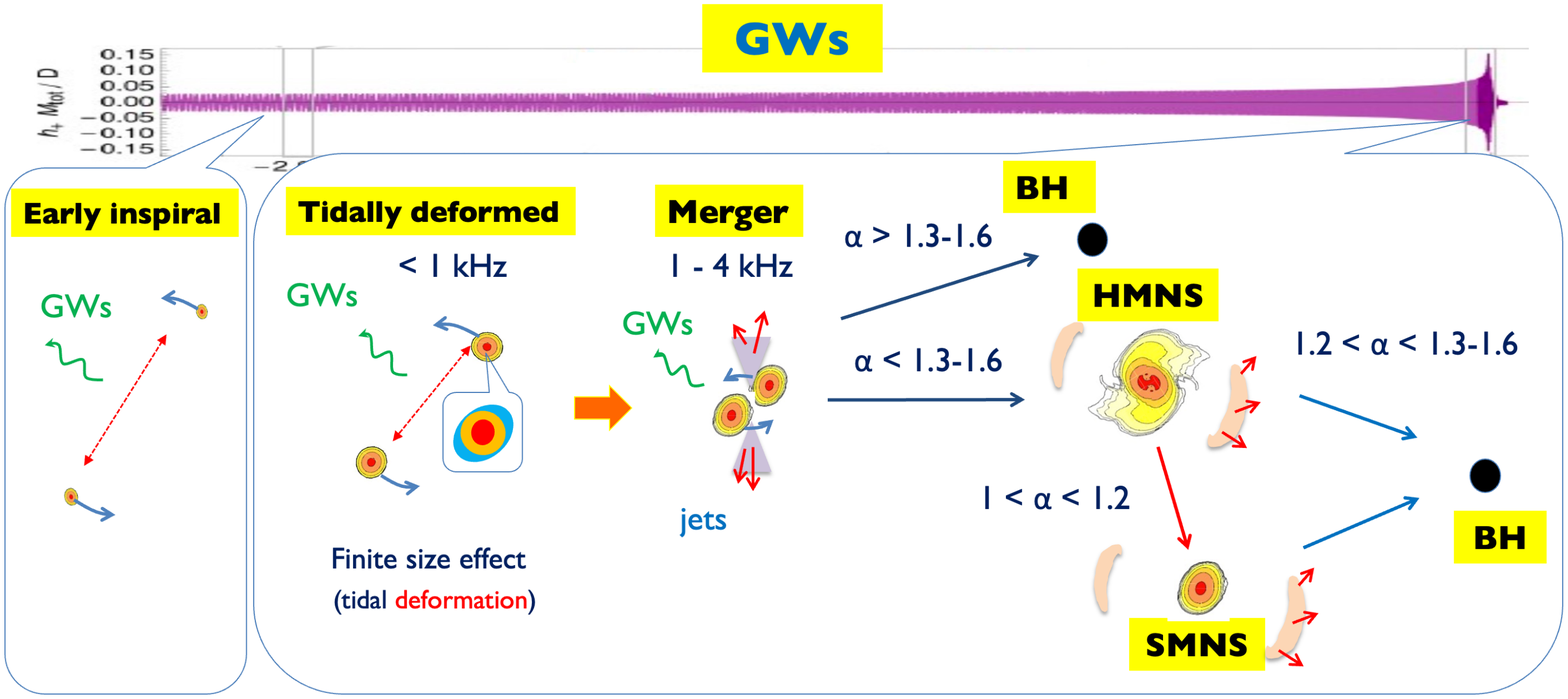}
\vspace{-0.8cm}
\caption{ The time evolution in a neutron star merger event including the inspiral, tidally deformed, and post-merger phases.
The fate after neutron star collisions depend on the ratio of the remnant mass to the maximum mass of the static neutron star, $\alpha \equiv M/M_{\rm static}^{\rm max}$.
\label{fig:NS_mergers}
}
\end{figure*}

\begin{figure*}[ht]
\centering
\includegraphics[width=0.99\textwidth]{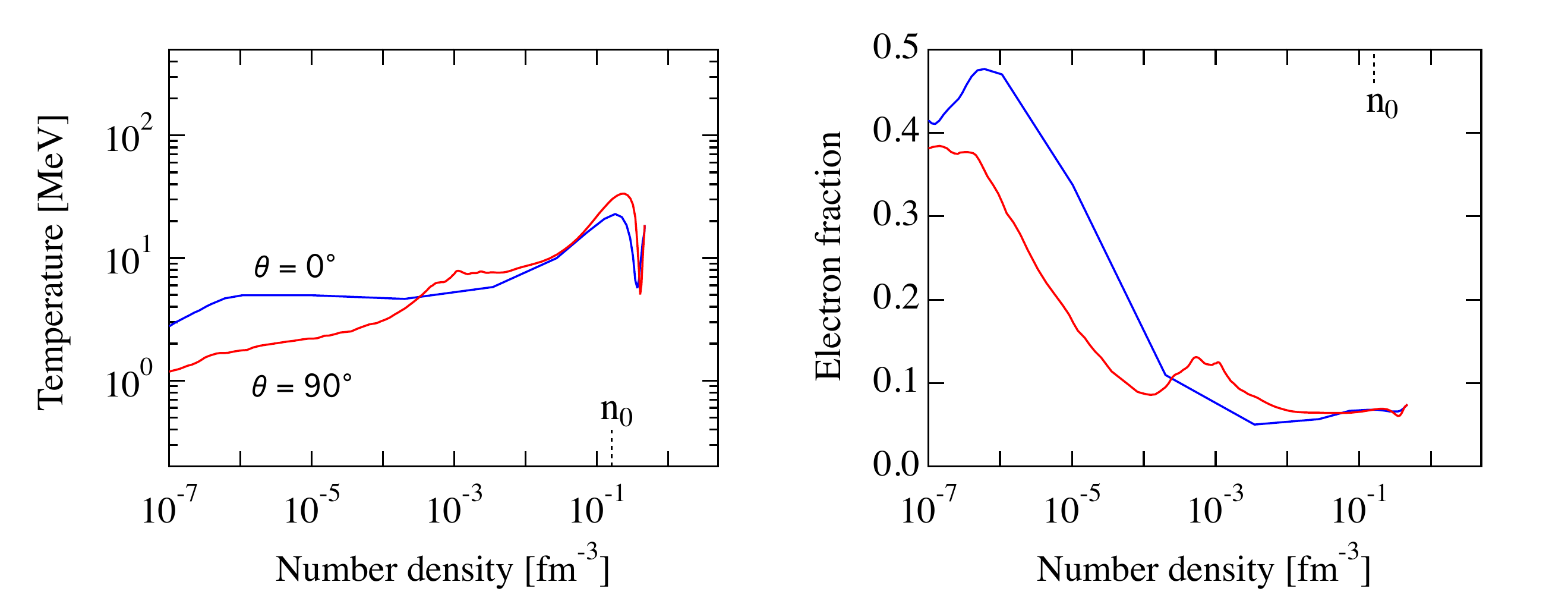}
\caption{Temperature and electron fraction of matter in a remnant of neutron star merger as a function of number density.  
The data are taken from the initial condition of the merger remnant in the numerical simulation under the axial symmetry \citep{fuj17,sum2021}.  
The conditions along the Z-axis ($\theta=0^{\circ}$) and the equatorial plane ($\theta=90^{\circ}$) are plotted by blue and red lines, respectively \citep{2017JPhCS.861a2028S}.  
\label{fig:NSMerger_Z_R}}
\end{figure*}

A merger of a neutron star binary offers a lot of information of neutron stars, 
from the static properties of cold neutron stars to dynamical processes of warm and dense matter after the collisions.
Fig.\ref{fig:NS_mergers} displays the time evolution.
There are a variety of signals including gravitational waves, electromagnetic waves, and neutrinos,
and hence neutron mergers are important targets in the {\it multi-messenger} astronomy.
So far gravitational waves have been detected in several events (including a blackhole) since 2015.
As for binary neutron star mergers, the clearest is GW170817 with the total mass of 
$M_{\rm tot} \simeq 2.74^{+0.04}_{-0.01} M_\odot$ detected by aLIGO and Virgo \citep{LIGOScientific:2017vwq}; 
in this event gravitational waves at frequencies $< \sim 1$kHz and electromagnetic counterparts
associated with the neutron star collisions have been detected.
The distance to the event was shorter ($43.8^{+2.9}_{-6.9}$ Mpc) than the other candidates of merger events for which electromagnetic signals were too faint.
Below some basic aspects of neutron mergers are discussed taking up GW170817 as an example.

Neutron star mergers include long time histories of neutron star binaries:
initially two neutron stars are widely separated, and each neutron star can be treated as if a point particle.
This is called inspiral phase which may last for million years or even longer. 
From the binary motion one can at least constrain the chirp mass, a combination of masses of two neutron stars,
or extract both neutron star masses if there are other additional information.
The binary slowly loses the energy by emitting gravitational waves, 
since the binary behaves as a time dependent quadrupole source of gravitational fields \citep{Hinderer:2009ca}.

After long time, the considerable orbital energy is lost, and the two neutron stars come close together.
Here the finite size effects of each neutron star set in;
neutron stars get deformed.
This stage is called tidally deformed phase whose characteristic time scale is the order of miliseconds.
The tidal deformation of each neutron star adds gravitational fields which increase the gravitational attraction between two neutron stars.
Accordingly, the merging process is accelerated and it increases the gravitational wave oscillating frequencies to a few hundred Hz.
The tidal deformability of each neutron star is highly sensitive to the compactness; a larger deformability for a larger radius.
The analyses of GW170817 in the tidal deformed phase put the upperbound on the tidal deformability and hence on the radius.
It is $R_{1.4} \lesssim 13.4$ km for $1.4M_\odot$ neutron stars \citep{Annala2018,Most:2018hfd}.

Eventually neutron stars contact and form a very massive object whose mass usually exceeds the maximum mass of static neutron stars,
and hence can survive only for a short time before collapsing to a black hole.
The time scale for such a collapse depends on many factors, including equations of state, 
rotational effects, finite temperature effects, viscous effects, magnetic fields, and so on \citep{Ravi:2014gxa,Shibata:2019wef}. 
It also depends on how neutron stars contact.
If the mass ratio is large, the contact is more head-on; the lighter one is dragged by the heavier one.
More typical are events where both neutron stars have the masses of $\sim 1.4M_\odot$.
In this case the contact is not head-on but with a large impact parameter. 
The collisional interface with density of $\sim 1-2n_0$ is heated to $\sim 40$ MeV,
while the core of each neutron star remains rather cool with the temperature of $5-20$ MeV.
The core of the merger reaches $\sim 4-7n_0$.
Some examples are shown in Fig.\ref{fig:NSMerger_Z_R}.

Unless the merger promptly collapses to a black hole (typical time scale is a millisecond), 
neutron stars have enough time to produce dynamical neutron rich ejecta of $\sim 10^{-2} M_\odot$ with $Y_p\sim 0.1-0.3$. 
This neutron rich condition differentiates neutron star mergers from supernovae with $Y_p\sim 0.3-0.4$.
In the ejected neutron rich matter, the rapid neutron capture (r-process) by nuclei occurs more quickly than the $\beta$-decay process,
and atomic number can grow into $A \sim 250$ overcoming moderate Coulomb repulsion in neutron rich nuclei.
The electromagnetic counterparts of the GW170817 event deliver the signals of the r-process and 
confirmed the expectation that the neutron star mergers can offer the place for the r-process \citep{Kasen:2017sxr}.
From the amount of the ejecta, the prompt collapse of the merger in GW170817 seems unlikely.
Some researches further used the amount of the ejecta to constrain the compactness of neutron stars \citep{Bauswein:2017vtn,Radice:2018ozg}.

If the merger is too massive, the prompt collapse prevents materials from flying away.
The observed ejecta in GW170817 support the scenario of non-prompt collapse, requiring sufficient stiffness for high density equations of state,
and disfavoring neutron stars with too small radii.
How long the merger oscillates strongly depend on the rotation that effectively increases the maximum mass.
It is known that a rigidly rotating neutron star has the maximum mass enhanced from the static counterpart 
by $\simeq 20$\%, $M^{\rm max}_{\rm rigid} \simeq 1.2 M^{\rm max}_{\rm static}$ \citep{Breu:2016ufb}. 
If the rotation is even faster, then rotational speed differs from the surface and interior.
Such a differentially rotating neutron star is metastable and has the mass higher than the static counterpart by $ 30-60$\%,
$M^{\rm max}_{\rm diff} \sim 1.3-1.6 M^{\rm max}_{\rm static}$ \citep{Baiotti:2016qnr}.
Neutron stars with the masses $M$ in the range of $M^{\rm max}_{\rm static} < M \lesssim 1.2M^{\rm max}_{\rm static}$ are called supermassive neutron stars (SMNS),
while in $1.2 M^{\rm max}_{\rm static} \lesssim M \lesssim 1.3-1.6M^{\rm max}_{\rm static}$ are called hypermassive neutron stars (HMNS). 
Seminal works assumed the GW170817 merger mass of $M\simeq 2.73-2.78M_\odot$ in the HMNS window, 
leading to the constraint $1.71-2.14M_\odot < M^{\rm max}_{\rm static} < 2.28-2.32M_\odot$  
\citep{Margalit:2017dij,Ruiz:2017due,Rezzolla:2017aly,Shibata:2019ctb}. 
Meanwhile the work assuming the SMNS scenario leads to the constraint $M^{\rm max}_{\rm static}  \gtrsim 2.3M_\odot$ \citep{Yu:2017syg}. 
It is important to examine which scenario is realized;
the prompt collapse is $\sim 1$ms, the life time of HMNS $\sim 0.01-1$s, and the life time of SMNS from $\sim10$s to several hours.
But the collapse time has not been measured.
For now the gravitational waves in GW170817 are not detected for the post-merger phase 
in which the frequency is $ \gtrsim 1$kHz; in such high frequencies the current detectors cannot differentiate signals from noises,
see the current status for the upgrade [https://dcc.ligo.org/LIGO-T1800044/public].

There are several additional observables which remain to be seen.
Neutrinos from neutron star mergers as well as gravitational wave signals in the post-merger phase have not been observed. 
Neutrinos should carry information of temperatures and matter composition.
The gravitational waves after a merger allow us to differentiate the SMNS and HMNS as a remnant of the merger and constrain equations of state \citep{Takami:2014zpa}.
Also, high frequency gravitational waves after mergers may be used to study the presence of first order phase transitions
which are usually discussed in the context of phase transitions from hadronic to quark matter.
They would also reveal significant enhancement 
in frequency due to radical shrinkage of the radius associated with softening of equations of state \citep{Most:2018eaw,Bauswein:2018bma,Blacker:2020nlq}.
The hadron-to-quark crossover scenarios are also being studied recently \citep{Huang:2022mqp,Fujimoto:2022xhv}.
 

\section{\textit{Quarks in neutron star matter}}

Neutron stars with the masses greater than $2M_\odot$ may accommodate matters with the densities beyond the purely hadronic regime.
Below quark matter is discussed as a candidate of matters beyond the hadronic regime.

\section{\textit{Free quark gas}}

\begin{figure}[tbh]
\vspace{-1.8cm}
\begin{center}	
\includegraphics[width=10.2cm]{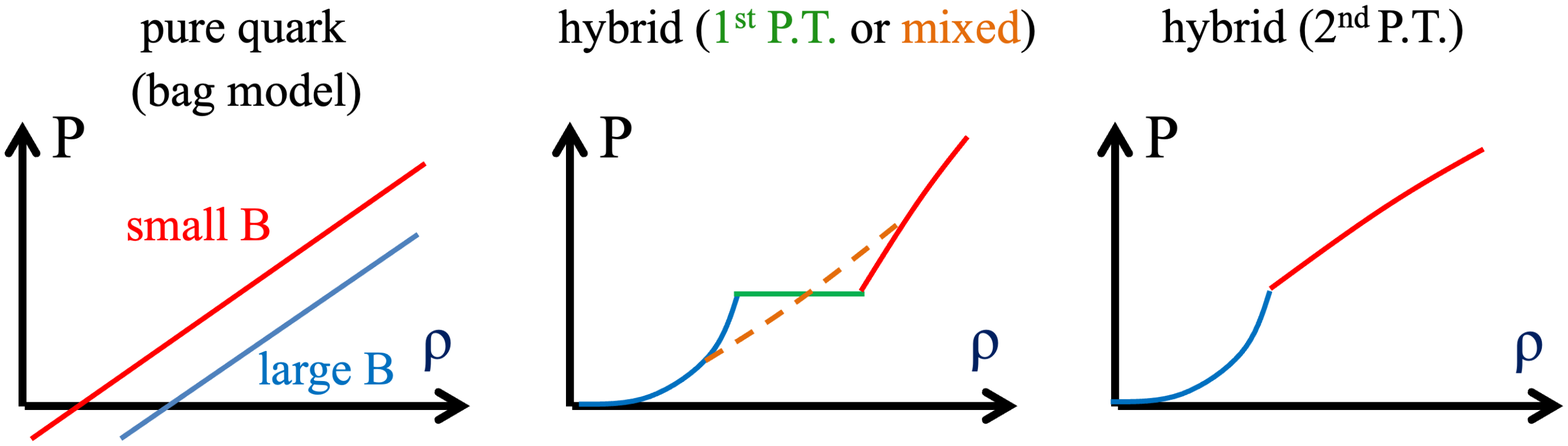}
\vspace{-1.5cm}
\caption{ $P$ v.s. $\rho$ in various types of models: (Left) A pure quark equation of state, e.g., a bag model. (Middle) A hybrid equation of state with first order phase transitions. 
A mixed phase based on the Gibbs construction is shown in the dashed line.
(Right) A model with a second order phase transition in which $\rmd P/\rmd \rho$ changes discontinuously.
}
 \label{fig:PT_types}
\end{center}		
\vspace{-0.5cm}		
\end{figure}

The simplest type of quark equations of state is an ideal gas with deconfined quarks. 
In the limit of massless quarks ($\mu_q = \mu/N_c$), the pressure is
\beq
P(\mu) = \frac{\, N_f  N_c \,}{\, 12 \pi^2 \,} \mu_q^4 - B \,,
\eeq
%
where $B$ is called a bag constant.
The baryon number and energy densities are computed as
(reminder: the thermodynamic relation $\rho = \mu n - P$)
\beq
n = \frac{\, \partial P \,}{\, \partial \mu\,} = \frac{\, N_f \,}{\, 3\pi^2 \,} \mu_q^3\,,~~~~~~~~~~
\rho = \frac{\, N_f N_c \,}{\, 4\pi^2 \,} \mu_q^4 + B  = 3P + 4B \,.
\eeq
The (adiabatic) sound velocity of matter is
\beq
c_s^2 = \frac{\, \partial P \,}{\, \partial \rho \,} = \frac{\,1\,}{\, 3 \,} \,.
\eeq
The causality requires the sound velocity to be smaller than the light velocity, and the ideal quark gas satisfies this constraint.
But the value $1/3$, called the conformal value, is actually quite large compared to those in terrestrial matters, reflecting the relativistic nature of the massless quark equations of state.
The equations of state of relativistic particles are naturally stiff since large pressure is developed by particles whose kinetic energy dominates over the rest mass energy.
This simple picture of quark matter, however, is made complicated by a bag constant which increases the energy while decreases the pressure (and thus softens equations of state).
A schematic plot is shown in Fig. \ref{fig:PT_types}.

\subsection{\textit{Quark star}}

At very high density, quark matter descriptions should become more effective than hadronic ones. 
If one trusts quark matter descriptions down to the quark chemical potential $\mu_c$ such that $P(\mu= N_c \mu_c)=0$, 
the corresponding baryon density, $n=n_q/N_c$, is nonzero,
\beq
n_c 
= \frac{\, N_f \,}{\, 3\pi^2 \,} \mu_c^3 
\simeq 1.78 n_0 \times
\bigg( \frac{\, N_f \,}{\, 3 \,} \bigg)^{1/4} 
\bigg( \frac{\, B^{1/4} \,}{\, 146\, {\rm MeV} \,} \bigg)^{3}
\,.
\eeq
The finite number density at zero pressure means that the matter is self-bound.
A chunk of such a self-bound quark matter can form a quark star with the arbitrarily small mass.
This should be contrasted to ordinary compact stars 
which require macroscopic amount of matter as a source of a strong gravity to trap neutron dominated matter (which is not self-bound).
Witten calculated the $M$-$R$ relations for quark stars and found the scaling for the maximum mass and the corresponding radius
\citep{Witten:1984rs},
\beq
M_{max} \simeq 2.0 M_\odot \times \bigg( \frac{\, 1.78 n_0 \,}{\, n_c \,} \bigg)^{2/3}  \,,~~~~~~
R |_{ M_{max} } \simeq 10.7\,{\rm km} \times \frac{\, M_{\rm max} \,}{\, 2M_\odot \,} \,,
\eeq
for $N_f=3$.
The maximum mass can become arbitrarily large when $B$ or $n_c$ is reduced to smaller values.
Thus, even without interactions, a relativistic quark gas can be very stiff, leading to a very large maximal mass.
But the above estimate requires some precautions concerning the validity of pure quark matter descriptions.
A small value of $B$ leads to a small density at $\mu_c$.
Demanding the maximum mass of quark stars to be bigger than $2M_\odot$, 
the estimate $n_c < 1.78 n_0$ follows; presumably the matter is too dilute to entirely neglect the color confinement in QCD. 
Including the mass corrections makes the situation even worse by lowering $n_c$.

\section{\textit{Hybrid hadron-quark equations of state: first or second order phase transitions}}

In a very dilute regime pure quark matter descriptions are invalid because the color confinement does not allow quarks delocalized. 
The natural description is to use hadronic models in dilute regime and to switch to quark matter models at some density. 
Most typically, a transition from hadronic to quark matter is characterized by first order phase transitions.
To determine which phases are realized,  hadronic and quark matter pressures, $P_h (\mu)$ and $P_q (\mu)$, respectively, are compared at a given $\mu$.
The phase with the larger pressure is realized as the ground state. 
At the phase transition point $\mu=\mu_c$,
\beq
P_h (\mu_c) = P_q (\mu_c)\,,~~~~~~~ \frac{\, \partial P_h \,}{\, \partial \mu \,} \bigg|_{\mu_c} < \frac{\, \partial P_q \,}{\, \partial \mu \,} \bigg|_{\mu_c}  \,,~~~~~~({\rm first~order})
\eeq
where the pressure is equal but $n=\partial P/\partial \mu$ (and hence $\rho$) increases discontinuously. 
The discontinuous change is often regarded as the signature of deconfinement.
At the first order phase transition, the pressure is continuous but $\rho$ jumps, leading to the vanishing sound velocity, $c_s =0$.

Another possibility is the second order phase transitions where $P$ and $n$ are continuous
but the second derivative $\chi = \partial^2 P/(\partial \mu)^2$ jumps, $\chi_h < \chi_q$.
Accordingly the sound velocity reduces discontinuously at the phase transition. This can be seen from the expression,
\beq
c_s^2 = \frac{\, n \rmd \mu \,}{\, \mu \rmd n \,} =  \frac{\, n \,}{\, \mu \chi \,} \,,
\eeq
with which $(c_s^h)^2 > (c_s^q)^2$; the sound velocity is smaller in quark matter.

\section{\textit{Hadron-quark mixed phases}}

\begin{figure}[tbh]
\vspace{-1.2cm}
\begin{center}	
\includegraphics[width=10.2cm]{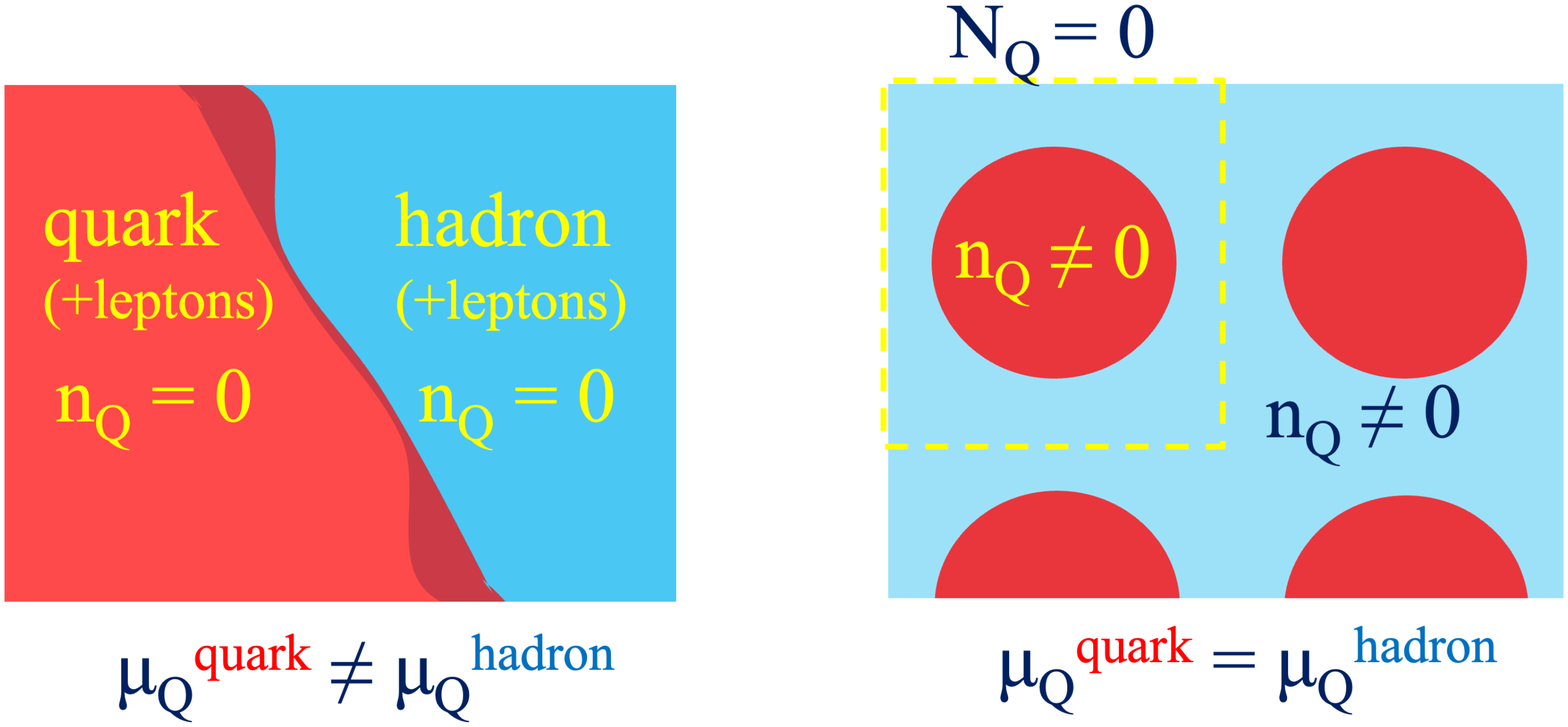}
\vspace{-1.cm}
\caption{ Maxwell (Left) v.s. Gibbs (Right) construction of hybrid equations of state.
}
 \label{fig:maxwell_gibbs}
\end{center}		
\vspace{-0.5cm}		
\end{figure}

Another possibility is a mixed phase of hadronic and quark matters
where the matter contains non-uniform domains of various kinds (pastas) \citep{Glendenning} (Fig.\ref{fig:maxwell_gibbs}).
To characterize such phases we first recall that neutron star matter contains baryon number and charge densities as conserved quantities.
Hence the pressure may contain two chemical potentials, $P(\mu,\mu_Q)$.
Imposing the neutrality condition $n_Q=0$, 
the charge chemical potential $\mu_Q$ is determined as $\mu_Q(\mu)$, 
so the pressure is written as $P(\mu) $.
In discussions of first order phase transitions in the last section,
$P_h (\mu)$ and $P_q (\mu)$ are compared assuming the conditions $n_Q^{hadron} =0$ and $ n_Q^{quark} =0$,
i.e., each phase is separately charge neutral.
Then, $\mu_Q^{hadron} (\mu) \neq \mu_Q^{quark} (\mu)$ in general.

In contrast, in mixed phases, the chemical equilibrium conditions are imposed for all components,
$\mu^{hadron} = \mu^{quark}$ and $\mu_Q^{hadron} = \mu_Q^{quark} $.
As a consequence, hadronic and quark matter are not separately charge neutral, but only the mixture is charge neutral.
This is achieved as $n_Q = \chi n_Q^{hadron} (\mu) + (1-\chi) n_Q^{quark} (\mu) = 0$ where $\chi$ characterizes the volume fraction.
The charge density has non-uniform distributions in space.

At low density, the finite size domains of a quark matter emerge in a hadronic matter, 
and the domain grows with density changing the shapes \citep{Alford:2001zr,Ju:2021hoy,Maruyama:2007ey,Nakazato:2008su}.
The quark matter is differentiated from hadronic domains by sharp surfaces, and the finite size domains are located in a periodic way.
Eventually the quark matter domains becomes bigger than the hadronic ones; the latters are immersed in a quark matter.
Eventually the hadronic domain disappears and the system is entirely described by a pure quark matter.
In this non-uniform descriptions, the $P$ vs $\rho$ relations do not contain jumps in $\rho$,
as the volume fraction $\chi$ interpolates pure hadronic and quark matter.
The sound velocity decreases in the interval of mixed phases.

\section{\textit{Hybrid hadron-quark equations of state: crossover scenario}}

\begin{figure}[tbh]
\vspace{-1.8cm}
\begin{center}	
\includegraphics[width=10.2cm]{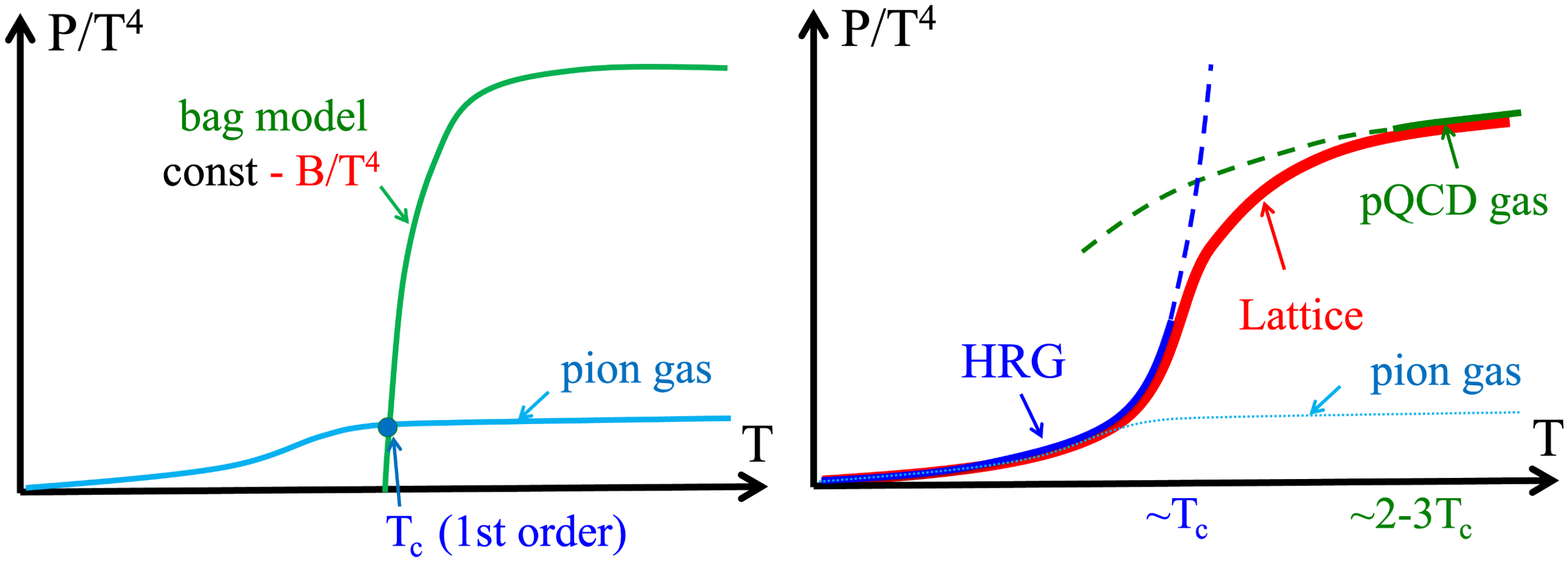}
\vspace{-1.5cm}
\caption{ A schematic description of normalized pressure, $P/T^4$, in hadronic and QGP phases. 
(Left) A pion gas v.s. a bag model for a QGP. The transition temperature is determined from the crossing point of two pressure curves. 
(Right) A more realistic comparison, HRG v.s. pQCD gas, together with the lattice result.
HRG and pQCD are consistent with the lattice results at low temperature ($T\lesssim T_c$)and high temperature ($T \gtrsim 2T_c$), respectively, but neither of them describe the intermediate region.
}
 \label{fig:HRG-QGP-trans}
\end{center}		
\vspace{-0.3cm}		
\end{figure}

A transition from hadronic to quark matter can be also a crossover with the derivatives of $P(\mu)$ being continuous.
To examine this possibility, it is instructive to discuss a finite temperature equation of state (at $\mu=0$) which includes a transition from a hadronic matter to a quark-gluon-plasma (QGP).
Lattice Monte-Carlo simulations, first principle calculations of QCD, 
have established that the transition is a crossover which begins to occur around $T_c \simeq 155$ MeV and continues to $T \sim 2$-$3T_c$ \citep{Aoki:2006we}.
At low temperature quarks and gluons are confined within hadrons and the system is dominated by a dilute hadron resonance gas (HRG). 
At higher temperature those hadrons are thermally excited and begin to overlap around $T\simeq T_c$. 
Then quarks and gluons should gradually become natural degrees of freedom to characterize the system.

Traditionally many works used a hybrid hadron-QGP model in which two phases are separated by a first order phase transition (Fig.\ref{fig:HRG-QGP-trans}).
The modeling is similar to the previous section. 
The simplest version of such hybrid models is to combine an ideal pion gas equation of state for hadronic matter $P^{\rm ideal}_{\pi}$ 
and a free quark and gluon gas with a bag constant for QGP equations of state.
Neglecting quark masses, for three flavors, ($d_{\rm QGP}=47.5$) \citep{yagi2005quark}
\beq
P^{\rm ideal}_{\rm QGP} (T) = d_{\rm QGP} \frac{\, T^4 \,}{\, 90 \,} - B \,,
\eeq
Again the bag constant is necessary to describe the phase transition; 
if $B=0$, $P^{\rm ideal}_{\rm QGP}$ would be always greater than $P^{\rm ideal}_{\pi}$ and no phase transition would happen.

Besides the orders of phase transitions,
it turned out that the above hybrid descriptions contain at least two problems. First, one cannot approach to $T\simeq T_c$ with the pion gas descriptions;
near $T_c$ equations of state are not dominated by pions but by other massive resonances, with the masses, $m_{\rm H}$, much greater than $T$.
Those massive contributions should be suppressed by a Boltzmann factor $\rme^{-m_{\rm H}/T}$, 
but the number of states grow fast with $T$ to compensate such suppression effects \citep{Hagedorn:1965st}.
Including resonances up to $E\sim 2.5$ GeV, the HRG model, $P_{\rm HRG}$, reproduces the lattice results up to $T\simeq T_c$ very well. 
Second, extrapolating the ideal gas model for QGP down to $\sim T_c$ substantially overestimates the pressure and entropy, 
as a consequence of neglecting confinement;
more realistically, as $T$ approaches $T_c$ from above, quarks and gluons should be trapped into hadrons with reduction of pressure and entropy.
The current perturbative QCD estimates of the pressure, $P_{\rm pQCD} $, seem valid down to $\sim 2T_c$ \citep{Ghiglieri:2020dpq}.

A description more conservative than direct comparisons of the extrapolated equations of state is to limit the use of HRG and QGP equations of state to the domain of applicability, 
and then to consider possible interpolations taking the hadronic equations of state at $\simeq T_c$ and QGP equations of state at $\simeq 2$-3$T_c$ as boundary conditions.
For the finite temperature case, one can simply take smooth curves for the interpolation.
For finite $\mu$ cases, the orders of transitions (first or second or crossover) are not established; 
case studies are necessary to prepare for future empirical determinations.

\section{\textit{Three window modeling} } 
 
\begin{figure}[tbh]
\vspace{-2.8cm}
\begin{center}	
\includegraphics[width=11.5cm]{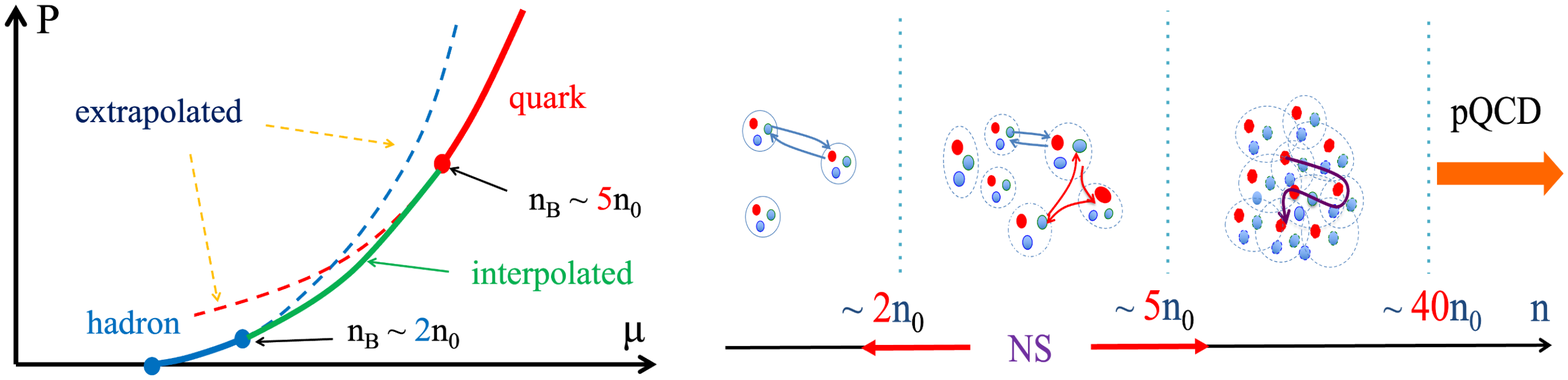}
\vspace{-2.2cm}
\caption{ A three-window modeling of a unified equation of state covering from nuclear to quark matter domains.
}
 \label{fig:3-window}
\end{center}		
\vspace{-0.3cm}		
\end{figure}

A three window modeling for dense QCD matter explicitly limits the domain of each model \citep{Masuda:2012kf,Masuda:2012ed}.
The three window refers to a nuclear (hadronic) matter at low density ($n \lesssim 2n_0$), quark matter at high density ($n \gtrsim 5n_0$), 
and a domain intermediate between the low and high density regimes ($ 2n_0 \lesssim n \lesssim 5n_0$).
In the following, first the physical picture is outlined and then a practical modeling is discussed.

In a nuclear matter regime, a matter is dilute and nucleons do not have many chances to interact.
Nuclear interactions are mediated by only few meson (or quark) exchanges.
Here one can use various nuclear equations of state, e.g. in Refs.\citep{Akmal:1998cf,Togashi:2017mjp,Steiner:2012rk,Typel:2009sy,Drischler:2021kxf}, 
which reproduce the nuclear laboratory experiments.

The dilute regime ends when many-body forces become sizable.
This is supposed to occur around $n \sim 2n_0$ \citep{Akmal:1998cf}.
A good measure to examine the convergence of many-body forces is the relative magnitudes of contributions from two- and three-body forces to the energy density.
For example, in case of contact interactions, 
the contributions from $N$-body forces to the energy density are supposed to grow as $\sim n^N$, increasing rapidly with $n$.
This raises questions whether nucleons remain reasonable effective degrees of freedom,
as baryon interactions are mediated by quark exchanges \citep{Fukushima:2020cmk}.
Meanwhile, the density is not high enough to trust quark matter descriptions as baryons do not largely overlap.
The problem here is the identification of proper degrees of freedom which is the starting point of any reliable calculations.

Beyond $\sim 5n_0$, baryons begin to overlap and our descriptions become simplified,
as quarks become the natural effective degrees of freedom.
This domain is regarded as quark matter,
irrespective of the existence of sharp hadron-quark phase transitions.
But quark matter in this regime is by no means weakly interacting.
The state-of-art pQCD calculations, including N$^2$LO and N$^3$LO soft contributions \citep{Gorda:2021znl,Gorda:2021kme},
have clarified that the domain of the weak coupling picture is $n \gtrsim 40n_0$.
In order to reconcile the quark pictures with such large $\alpha_s$ corrections,
presumably one must include non-perturbative effects to strongly renormalize physical parameters, 
e.g., effective masses and couplings in quark matter descriptions. 
Such strong renormalizations are rather common in hadron physics;
quark models for hadrons \citep{DeRujula:1975qlm,Hatsuda:1994pi}, 
reproducing many hadron properties,
are based on constituent quarks whose effective masses differ from those in the QCD Lagrangian.

As noted before,
the most theoretically challenging is the description of the intermediate regime between $\sim 2n_0$ and $\sim 5n_0$.
Fortunately, however, it is this domain where neutron stars of $M \simeq$ 1-2$M_\odot$ can give the most powerful constraints. 
The $M$-$R$ relations of neutron stars have the one-to-one correspondence with the equations of state
so that a better measurement of $M$-$R$ more precisely determines equations of state \citep{Lattimer:2000nx}.

\section{\textit{Three window model in practice} } 

One of possible ways to implement these pictures is to adopt the following phenomenological modeling \citep{Kojo:2014rca,Baym:2017whm}.
The first step is to choose the densities $n_L$ and $n_U$ at which nuclear and quark matter descriptions, respectively, are terminated.
For instance, $n_L =2n_0$ and $n_U=5n_0$.
Then, nuclear equations of state at $n \le n_L $ and quark equations of state at $n \ge n_U $ are prepared.
In practice, one may choose nuclear equations of state based on the modern nuclear forces and many-body calculations.
Meanwhile quark equations of state must be calculated by using some constituent quark models developed for hadron spectroscopy.
Finally, the nuclear and quark equations of state are phenomenologically interpolated.
For interpolating functions, one can choose, for instance \citep{Kojo:2014rca,Baym:2017whm}, 
(see also, e.g., \cite{Ayriyan:2021prr})
\beq
P_{\rm inter} (\mu) = \sum_{j=0}^5 c_j \mu^j \,,
\eeq
where the coefficients of the polynomials are fixed by demanding matching conditions up to the second derivatives of $P$,
\beq
 \frac{\, \rmd^l P_{\rm inter} \,}{\, (\rmd \mu)^l \,} \bigg|_{ \mu_{L} } 
=
\frac{\, \rmd^l P_{h} \,}{\, (\rmd \mu)^l \,} \bigg|_{ \mu_{L} } 
\,,~~~~~
\frac{\, \rmd^l P_{\rm inter} \,}{\, (\rmd \mu)^l \,} \bigg|_{ \mu_{U} } 
=
\frac{\, \rmd^l P_{q} \,}{\, (\rmd \mu)^l \,} \bigg|_{ \mu_{U} } \,,~~~(l=0,1,2)\,.
\eeq
The $\mu_L$ and $\mu_U$ are extracted from the conditions $n(\mu_L) = n_L$ and $n(\mu_U) = n_U$, respectively.
Six boundary conditions are available to determine all the coefficients.

In the above-mentioned procedures, one can construct a unified equation of state. 
But not all of those equations of state are physical.
Physical equations of state must satisfy the conditions of the thermodynamic stability ($\chi = \rmd^2P/\rmd \mu^2 \ge 0$) 
and the causality ($c_s^2 \le 1$).
In order to make interpolated equations of state physical, a proper combination of nuclear and quark equations of state must be chosen \citep{Baym:2019iky,Kojo:2021wax}.
In general, the causality tends to be violated when a nuclear equation of state is softer and a quark equation of state is stiffer,
because such soft-to-stiff combination requires the pressure grows rapidly as a function of energy density,
accompanying a large slope, $\rmd P/\rmd \rho = c_s^2$.
In the context of neutron stars physics, the nuclear equation of state up to $\sim 2n_0$ is strongly correlated with the radii of $\sim 1.4M_\odot$ neutron stars \citep{Lattimer:2000nx},
while the quark equation of state is correlated with the maximum mass of neutron stars which must be larger than $2M_\odot$.

If one assumes the first order phase transitions in the interpolated domain, 
it becomes more difficult to satisfy the causality constraint.
During the first order, $P$ is constant and $\rho$ grows. 
After the phase transition is over, the $\rmd P/\rmd \rho$ must grow even more steeply to achieve the stiffness 
necessary to pass the $2M_\odot$ constraints.
Of course, weak first order transitions are still possible
but such small transitions may be treated as a small perturbation to the crossover scenarios.
Thus, the three window model made of smooth interpolating functions may be taken as a baseline
to discuss more detailed phase structures of matter.

\section{\textit{Interactions in strongly correlated quark matter}}

\begin{figure}[tbh]
\vspace{-1.2cm}
\begin{center}	
\includegraphics[width=9.5cm]{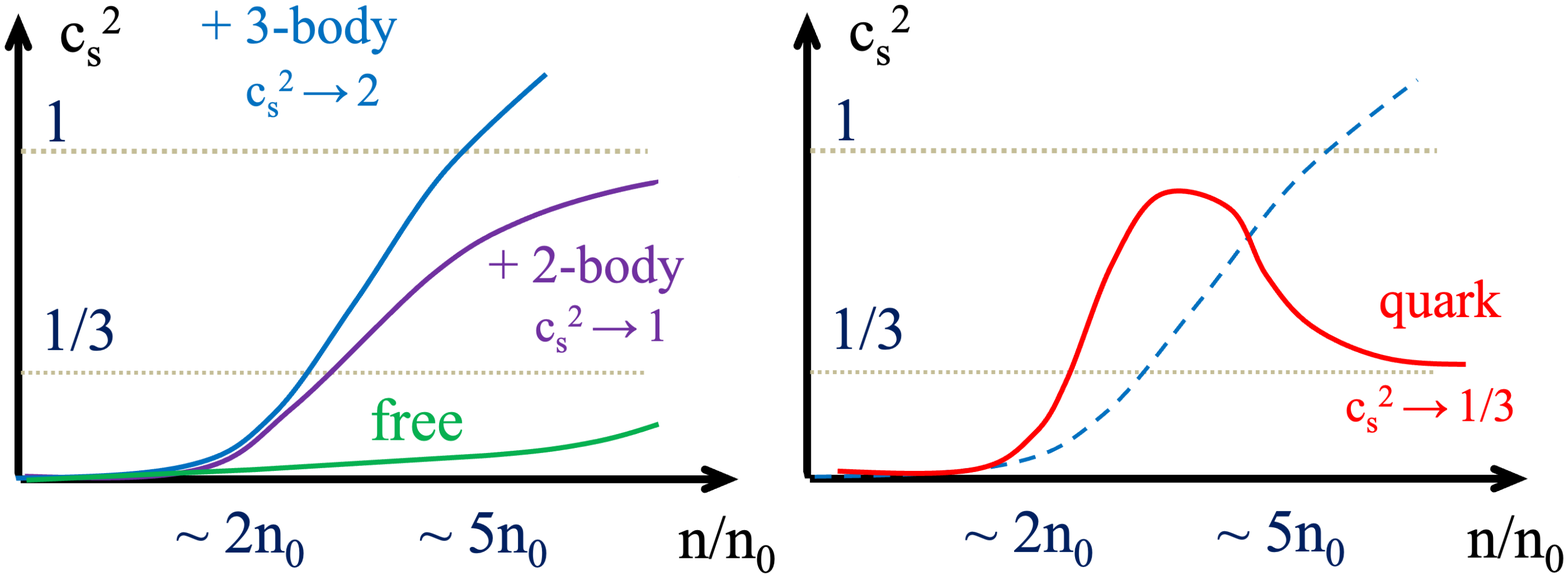}
\vspace{-1.2cm}
\caption{ A squared sound velocity $c_s^2$ as a function of $n/n_0$.
(Left) Purely nucleonic descriptions for free nucleons, plus two-body repulsion, and plus three-body repulsion. The repulsions are assumed to be the contact type.
(Right) The behaviors typical in crossover models arranged to satisfy nuclear and $2M_\odot$ constraints. Stiffening is typically more rapid than in purely nucleonic models.
}
 \label{fig:cs2}
\end{center}		
\vspace{-0.3cm}		
\end{figure}

The question is what kind of quark matter can generate equations of state which are stiff enough to pass the $2M_\odot$ constraints.
As noted before, non-interacting massless quarks can give a very stiff equation of state,
so starting with a free relativistic quark equation of state and then adding masses and interactions as corrections 
may be a good strategy. 

Before moving to the quark cases, it is useful to address how the extrapolation of a purely nucleonic model would give a stiff equation of state.
Here the following parametrization is considered,
\beq
\rho (n) = m_N n + a \frac{\, n^{5/3} \,}{\, m_N \,} + b n^\alpha \,,
\eeq
where the first is the mass density, the second kinetic energy, and the third is an interaction with $a,b$ positive constants.
Noting $\mu = \partial \rho/\partial n$ and $P=\mu n - \rho$, we find
\beq
P = \frac{\, 2 \,}{\, 3 \,} a \frac{\, n^{5/3} \,}{\, m_N \,} + b (\alpha-1) n^\alpha \,.
\eeq
The $2M_\odot$ constraint requires $P$ and $\rho$ to be the same order.
This is never achieved unless we have very strong repulsions;
for $b=0$, $P/\rho \sim n^{2/3} /m_N^2$, very small unless $n^{1/3} \sim m_N$ or $n\sim 50n_0$.
This regime is not achievable in neutron star cores.
Thus in the following we neglect the kinetic energy term. 
Next, it is instructive to assume $\alpha > 1$ and consider the regime where the interaction terms dominate over the rest mass energy.
Then
\beq
P \sim (\alpha-1) \rho ~\rightarrow ~ c_s^2 \sim (\alpha-1) \,.
\eeq
For contact interactions, the dominance of two-body forces with $\alpha \sim 2$ leads to $c_s^2 \sim 1$, 
while three-body forces with $\alpha \sim 3$ leads to $c_s^2 \sim 2$ with violation of the causality.
This trend indicates that, while many-body repulsions  help equations of state to achieve necessary stiffness, 
their dominance in equations of state would violate the causality.
As density increases, terms with the highest powers of $n$ dominate over the others
so that one eventually has to stop using models with more than two-body forces,
or must assume the couplings to strongly decrease at high density.
For example, time-honored Akmal-Phandhari-Ravenhall (APR) equation of state violates the causality at $\sim 5n_0$ before reaching the maximum mass
\citep{Akmal:1998cf}.

For a relativistic quark matter \citep{Alford:2004pf}.
it is useful to consider a simple parametrization,
\beq
\rho (n) = a n^{4/3} + b n^\alpha \,,
\eeq
where the first is the relativistic kinetic energy and the second describes interactions or mass energies.
As before, one can calculate $\mu$ and use $P=\mu n - \rho $. 
It is possible to derive a useful expression by eliminating $an^{4/3}$ terms in favor of $\rho$ to reach \citep{Kojo:2014rca}
\beq
P = \frac{\, \rho \,}{3} + b \bigg( \alpha - \frac{\, 4 \,}{\, 3 \,} \bigg) n^\alpha \,,
\eeq
where $n$ is a function of $\rho$.
The $b=0$ case leads to the conformal limit, $P=\rho/3$.

It is important to note that the effects of interactions enter as the product of $b$ and $(\alpha-4/3)$;
hence, not only the sign of $b$ but also the density dependence is important to judge whether interactions stiffen or soften equations of state.
For $\alpha > 4/3$, the repulsive interactions ($b > 0 )$ stiffens equations of state; an example is a contact quark $N (\ge 2)$-body repulsion 
characterized by $b>0$ and $\alpha = N$.
It is typical to discuss stiffening based on repulsions, 
but it should be kept in mind that terms with $\alpha > 4/3$ should not be extrapolated to very high density,
as they would dominate over the kinetic energy and contradict with the asymptotic free nature of QCD at short distance.

In order to find stiffening terms that have the natural high density limit, one can consider $\alpha < 4/3$ where the attractive interactions ($b<0$) stiffens equations of state.
Terms with the powers of $n$ less than $4/3$ must accompany some mass scales other than $n$.
A possible term is the mass term from the expansion of the quark energies,
$E \sim p(1 + m_q^2/p^2 +\cdots)$ ($m_q$: quark mass), but it yields $\rho_{\rm mass} \sim + m_q^2 n^{2/3}$, softening equations of state.
The bag constant, with $b>0$ and $\alpha=0$, also softens equations of state.
Yet there are still other possibilities that the dynamical scale of QCD, $\Lambda_{\rm QCD} \sim 200$-$300$ MeV, appear in attractive correlations,
leading to an energy density $\rho_{\rm dyn} \sim - \Lambda_{\rm QCD}^2 n^{2/3}$.
This case stiffens equations of state.
The factor $n^{2/3}$ indicates that the desired term is related to the non-perturbative dynamics near the Fermi surface whose area is $\sim 4\pi p_F^2 \sim n^{2/3}$.
This brings our attention to the physics near the Fermi surface in quark matter.
Those include, e.g., pairing effects in color-superconductivity (CSC), baryonic correlations in quarkyonic matter, and so on, which will be discussed in the following.

\begin{figure}[tbh]
\vspace{-1.2cm}
\begin{center}	
\includegraphics[width=9.cm]{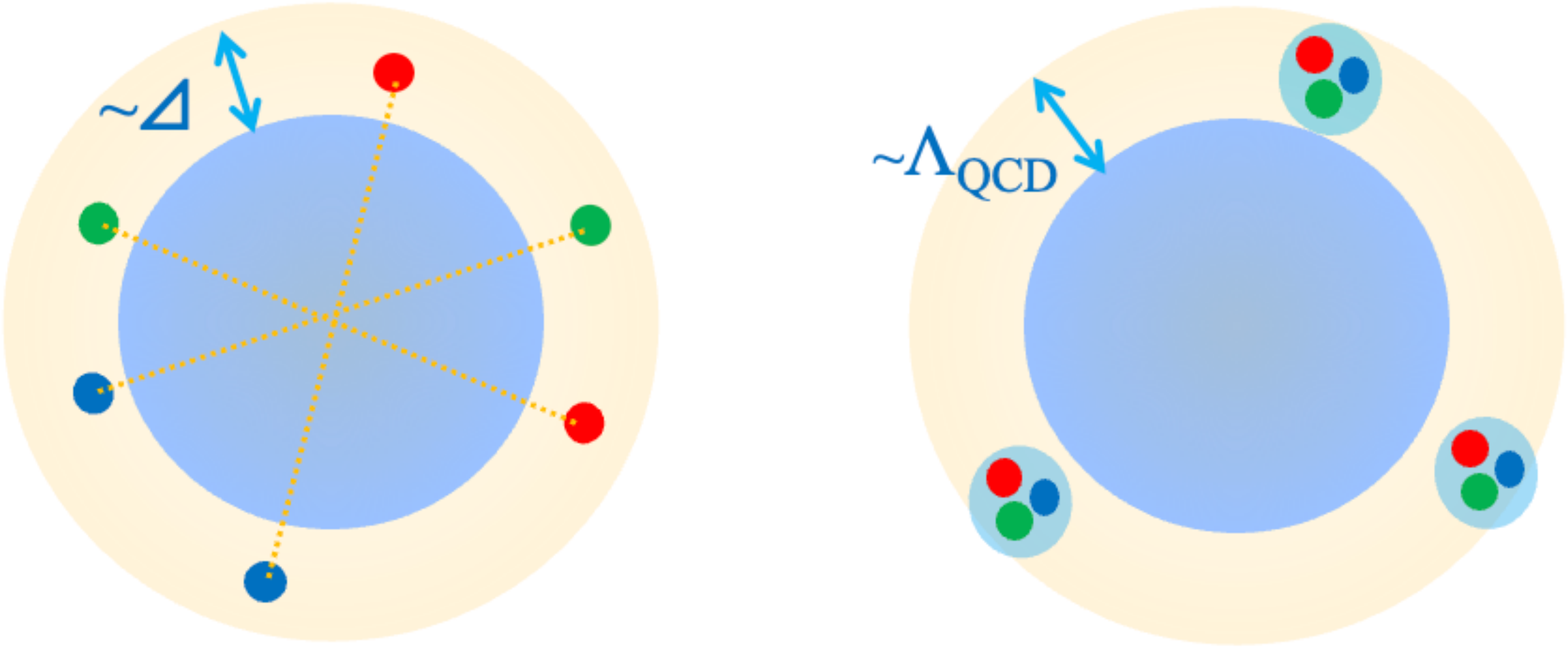}
\vspace{-1.2cm}
\caption{ The quark Fermi sea with non-perturbative correlations near the Fermi surface.
(Left) The diquark pairing in the S-wave, color-, spin-, and flavor-antisymmetric channels.
The Fermi surface is diffused over the momentum scale of the energy gap $\sim \Delta$.
(Right) Three-particle or baryonic correlations discussed in quarkyonic matter scenario.
}
 \label{fig:cs2}
\end{center}		
\vspace{-0.3cm}		
\end{figure}

\section{\textit{Diquark pairings in color-superconductivity (CSC)} }

The CSC is one of very popular scenarios in the high density limit.
Ref.\cite{Alford:2007xm} is recommended as an extensive review.
Here only the basic idea is mentioned.

The color-superconductivity is triggered by the condensations of quark-quark pairs (diquarks)
as those of electron-electron pairs in usual superconductivity \citep{schrieffer1999theory}.
The energetic costs to populate quarks near the Fermi sea are small,
and attractive diquark correlations trigger the macroscopic appearance (condensates) of diquark pairs.
They reorganize the ground state and open a gap $\Delta$ in a quark spectrum which reflects the energy required to excite a quark.
The appearance of the gaps drastically changes the thermal and transport properties of quark matter
since quarks cannot be activated by weak perturbations.

In QCD, gluon exchanges offer various attractive channels for quark-quark interactions, as quarks have color, flavor, spin, and orbital degrees of freedom.
As for colors, quarks (antiquarks) belong to the representation ${\bf 3}\, ({\bf \bar{3} }) $.
Diquarks can have the representations ${\bf 3} \otimes {\bf 3} = {\bf \bar{3} } \oplus {\bf 6}$, where the former (latter) is an anti-symmetrized (symmetrized) color state.
A diquark in ${\bf \bar{3} }$ has the smaller color charge than two separated quarks, producing less color-electric fields with smaller energetic costs.
Thus it is natural to consider ${\bf \bar{3}}$ channels.
For relative orbital wavefunctions, the S-wave channel can utilize the entire Fermi surface to make diquark pairs.
Now the Fermi statistics demands the leftover spin-flavor combinations to be symmetric.
To proceed further, it is important to consider the color-magnetic interactions, 
$ \sim - c ( \vec{\lambda}_1 \cdot \vec{\lambda}_2 ) ( \vec{\sigma}_1 \cdot \vec{\sigma}_2)$, arising from a one-gluon-exchange
with $c$ being positive, $\vec{\lambda}$ the Gell-Mann matrix for colors, and $\vec{\sigma}$ is the Pauli matrix for spins \citep{DeRujula:1975qlm}.
For the ${\bf \bar{3}}$ color states, $ \langle (\vec{\lambda}_1 \cdot \vec{\lambda}_2 ) \rangle_{ {\bf \bar{3}} } <0$,
so the spin should be singlet to get an energetic benefit.
Accordingly, the flavor states are fixed to antisymmetric states ${\bf \bar{3}}$ (${\bf 1}$) in three (two) flavor theories.
In three flavors, the phase with the above-mentioned diquark condensates is called {\it color-flavor-locked} (CFL) phase \citep{Alford:1998mk}.
The diquark contributions to the energy density is 
\beq
\rho_{\Delta} \sim - \Delta^2 (4\pi p_F^2) \sim  - \Delta^2 n^{2/3} \,.
\eeq
If $\Delta$ depends on $n$ only weakly, the diquark condensates stiffen equations of state.

In the high density limit, the magnitude and density dependence of $\Delta$ can be rigorously estimated within the weak coupling framework \citep{Son:1998uk} 
\beq
\Delta \sim \mu_q g_s^{-5} \exp{ \bigg(- \frac{\, 3\pi^2 \,}{\, \sqrt{2} g_s \,} \bigg) } \,.
\eeq
The gap $\Delta$ is essentially determined by two ingredients; (i) the strength of interactions; and (ii) the phase space available for quantum fluctuations
(or intermediate states appearing in the self-energy calculations).
The high momentum transfer processes, for which $\alpha_s$ is small, are weak in interactions but there are many intermediate states to enhance the magnitudes.
Meanwhile, small momentum transfer processes have stronger coupling but not much phase space is available. 
At high enough densities, the large momentum processes dominate, 
and the estimate of $\Delta$ can be done within the weak coupling regime.

Such high density estimate, however, is questionable in applications to the NS physics.
The Fermi surface is not large enough to be dominated by weak coupling processes;
a typical momentum transfer is $p_F < 1$ GeV and hence $\alpha_s$ is not small.
The estimates of $\Delta$ or the existence of the CSC at strong coupling is not established.
But one can gain some insights from two-color QCD, a cousin of our three-color QCD.
Unlike the three-color case, the lattice Monte-Carlo simulations at finite density
are doable in two-color QCD. Various quantities, e.g., phase structures and equations of state, have been computed \citep{Boz:2019enj,Iida:2019rah,Bornyakov:2022pdd}.
The lattice simulations showed the existence of diquark condensed phase as model calculations predicted.
What is quite remarkable is that, even at $n \gtrsim 10n_0$, the melting temperature of diquark condensates
is high, $T_c \simeq 100$ MeV and is insensitive to density at least to $\sim 50$-$100n_0$. 
In n\"aive BCS estimate, the $T_c$ is related to the gap as $\Delta \simeq 175$ MeV. 
The value is comparable to $\Lambda_{\rm QCD}$, suggesting the non-perturbative physics persist to high density
close to, or even above, the domain where pQCD has been trusted.

\section{\textit{Quarkyonic matter}}

In addition to diquarks one can also consider baryonic correlations near the Fermi surface.
This should be increasingly important in the lower density regime.
Diquarks, which have colors, cannot survive in dilute regime because of the confinement,
and must pick up a quark to get color neutralized.
This consideration draws our attention to the concept of a {\it quarkyonic matter} \citep{McLerran:2007qj}.

The quarkyonic matter was originally discussed in the limit of large number of colors (large $N_c$) 
with $g_s^2 N_c = O(1)$ fixed \citep{tHooft:1973alw}.
In this large $N_c$ limit, the fermion loops of $O(1/N_c)$ are negligible compared to the gluon loops,
so that the gluons are unaffected by the quark dynamics. 
This picture leads to an interesting consequence at high density.
As $\mu$ increases, the baryonic matter appears at $\mu_q = \mu/N_c \sim M_q \sim M_B/N_c$ where $M_q$ is the constituent quark mass.
Increasing $\mu_q$ further to $\mu_q \gg \Lambda_{\rm QCD}$, baryons eventually overlap and quarks form a Fermi sea.
This looks paradoxical in the large $N_c$ limit:
while baryons largely overlap, 
gluons must remain the same as in vacuum
and should confine any colored objects.
More precisely, the gluons are unaffected until the screening effects become strong enough to dominate over non-perturbative effects.
The Debye mass reaches the non-perturbative scale as
\beq
m_D^2 \sim g_s^2 \mu_q^2 \sim \Lambda_{\rm QCD}^2 
~\rightarrow~ \mu_q \sim N_c^{1/2} \Lambda_{\rm QCD} \,.
\eeq
Large $N_c$ emphasizes the possibility of an window $\Lambda_{\rm QCD} \ll \mu_q \ll N_c^{1/2} \Lambda_{\rm QCD}$,
where a seemingly paradoxical matter, {\it a quark matter with confinement}, is realized.

There are several scenarios to resolve the paradox.
The first is to just assume that baryons remain effective degrees of freedom, entirely hiding quarks from the dynamics
until the condition $\mu_q \sim N_c^{1/2} \Lambda_{\rm QCD}$ is met.
The constituents of the Fermi sea are color-singlet,
so is the resulting Fermi sea.
But the description is not necessarily natural; the kinetic energy $\sim p_F^2/m_N \sim 1/N_c$ is much smaller than the interaction energy $\sim N_c$ \citep{Witten:1979kh}.
This raises questions concerning the choice of proper degrees of freedom.
On the contrary, the quarkyonic matter scenario does not demand the Fermi sea to be made of confined objects, 
but demand only the Fermi sea as a whole to satisfy the color-singlet condition. 
In this picture, one may take quarks as natural degrees of freedom to describe the bulk part of the Fermi sea, as in conventional quark matter pictures at weak coupling.
Meanwhile, quarks near the Fermi surface are subject to soft gluon exchanges as the phase space for the final states is open. 
Such small momentum transfer does not change much the location of quarks in momentum space but the coupling is very strong and confining.
An efficient way to include such confining effects is
to choose baryons as effective degrees of freedom near the quark Fermi surface.
A quarkyonic matter describes a quark matter with a baryonic Fermi surface.

In the context of neutron star physics, the quarkyonic matter picture explains stiffening of matter in the transition from a nuclear to a quark matter \citep{McLerran:2018hbz,Jeong:2019lhv}.
The essential ingredient is the quark Pauli blocking constraint on baryonic degrees of freedom.
To see how it works, it is useful to consider the occupation probability of quark states, $f_q(p;n)$, at a given density \citep{Kojo:2021ugu}.
At very high density, it should describe a usual quark Fermi sea, $f_q \sim \theta(p_F-p)$.
The question is how it evolves from nuclear to quark matter.
In a single nucleon, quarks are localized within the domain of $\sim 1\,{\rm fm}^3$, 
and thus occupy momentum states from zero to $\sim \Lambda_{\rm QCD}$.  
The occupation probability for each state is much less than 1.
With such small occupation probability, the quark Pauli blocking is not important in a dilute baryonic matter.
But as baryon density increases, such small occupation probability from each baryon is accumulated,
and at $n\sim \Lambda_{\rm QCD}^{3}$, the low momentum states begin to get saturated, approaching to the usual $f_q \sim \theta(p_F-p)$ type distribution.
In this quark {\it saturation} regime, the quark Pauli blocking constraint must be taken into account in baryonic descriptions.

The above idea is naturally implemented in the quarkyonic matter description.
At low density there is no quark Fermi sea and only a baryonic Fermi surface is available.
At densities of the quark saturation, a quark Fermi sea is established, pushing the baryonic Fermi surface to the high momentum domain.
The resulting baryons, with quarks collectively moving in the same directions, have high momenta and are relativistic.
The pressure is naturally large.
This situation should be contrast to baryons in dilute regime where quarks do not have the common orientation;
the forces from quarks cancel within a baryon, without yielding much pressure.
Before and after the quark saturation,
the pressure increases rapidly, but the energy density changes only smoothly.
In fact, even pure nuclear and pure quark matter do not have much difference in the energy densities:
they have $ \rho \sim m_N n \sim N_c \Lambda_{\rm QCD} n$ and $\sim N_c n^{4/3}$ respectively,
which are the same order at $n\sim \Lambda_{\rm QCD}^3$.
This rapid increase in $P$ with gentle change in $\rho$ leads to a peak in sound velocity.

\begin{figure}[tbh]
\vspace{-1.2cm}
\begin{center}	
\includegraphics[width=12.cm]{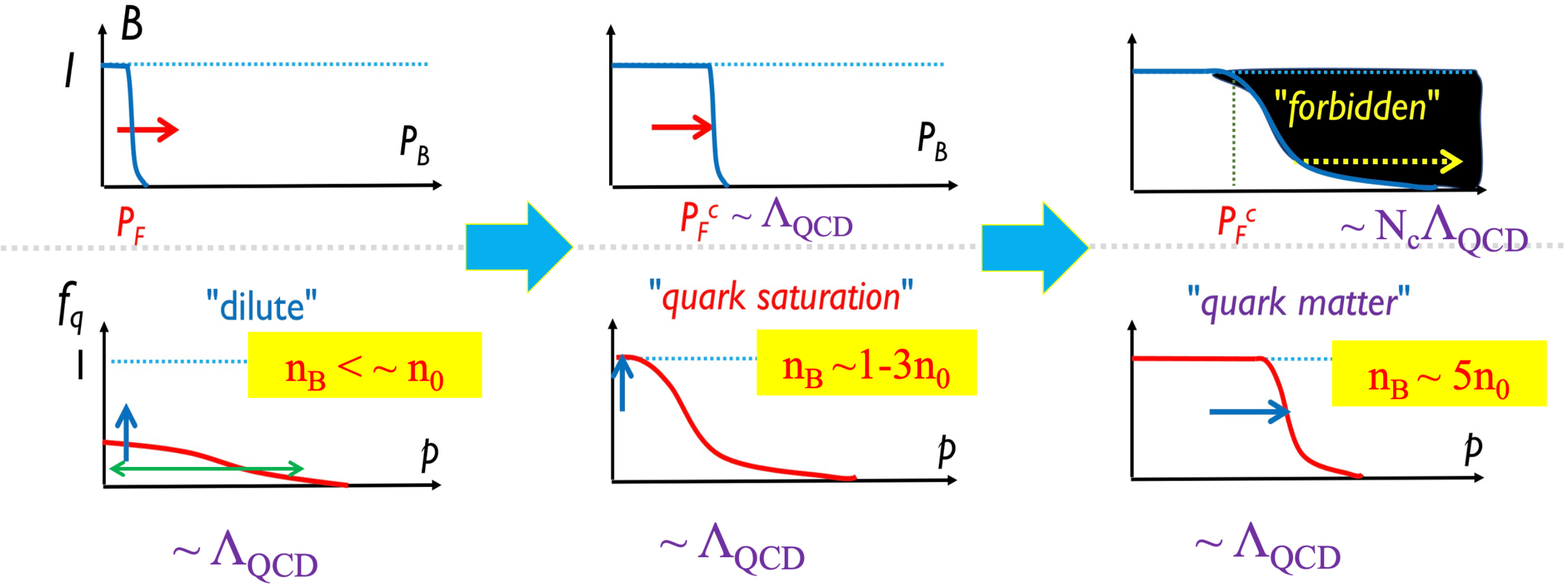}
\vspace{-1.8cm}
\caption{ The relation between the baryon ($B$) and quark ($f_q$) occupation probabilities for given momenta.
In dilute baryonic matter quarks are packed within a baryon and have broad distribution in momenta; none of momentum states are saturated and quark Pauli blocking effects are minor.
As baryon density increases, eventually quark states at low momenta are saturated.
Beyond this regime, baryons must occupy higher momenta but with small occupation probabilities not to violate the quark Pauli principle.
The quarks eventually establish conventional Fermi sea.
}
 \label{fig:cs2}
\end{center}		
\vspace{-0.3cm}		
\end{figure}

\bibliographystyle{apsrev4-1}
\bibliography{merged}{}

\end{document}